\newif\ifonecolumn
\onecolumntrue 
\ifonecolumn
  \documentclass[journal,draftcls,onecolumn, 12pt]{IEEEtran}
\else
  \documentclass[journal]{IEEEtran}
\fi

\usepackage[T1]{fontenc}

\usepackage{amsmath}
\usepackage{cite}
\usepackage{array}
\usepackage{stfloats}
\usepackage{url}
\usepackage{amsmath,amssymb}
\usepackage{epsfig,fancyhdr}
\usepackage{CJKutf8}
\usepackage{times}
\usepackage{graphicx}
\usepackage{multirow}
\usepackage{stfloats}
\usepackage{subfigure}
\usepackage{color}
\usepackage[dvipsnames]{xcolor}

\usepackage{bm}
\newcommand{\Fig}{Fig.$\,$}

\newtheorem{thm}{Theorem}

\newtheorem{defin}{Definition}
\newtheorem{lemma}[thm]{Lemma}

\newtheorem{eg}{Example}
\newtheorem{remark}{Remark}
\newtheorem{theo}{Theorem}
\begin{document}

\title{Two-Dimensional Golay Complementary Array Sets With Arbitrary Lengths for Omnidirectional MIMO Transmission}

\author{You-Qi Zhao
, Cheng-Yu~Pai
, Zhen-Ming Huang
, Zilong~Liu
,~\IEEEmembership{Senior Member,~IEEE}, and Chao-Yu~Chen
,~\IEEEmembership{Member,~IEEE}
\thanks{This work was supported by the Ministry of Science and Technology,
Taiwan, R.O.C.,  under Grant MOST 109--2628--E--006--008--MY3 and MOST 111--2218--E--305--002.
}
\thanks{You-Qi Zhao and C.-Y. Pai are with the Department of Engineering Science, National Cheng Kung University, Tainan 701, Taiwan, R.O.C. (e-mail: n98081505@gs.ncku.edu.tw).}
\thanks{Z.-M. Huang is with the Institute of Computer and Communication Engineering,  National Cheng Kung University, Tainan 701, Taiwan, R.O.C. (e-mail: n98101012@gs.ncku.edu.tw).}
\thanks{Zilong~Liu is with the School of Computer Science and Electronic Engineering, University of Essex, United Kingdom (e-mail: zilong.liu@essex.ac.uk).}
\thanks{C.-Y. Chen is with the Department of Electrical Engineering and the Institute of Computer and Communication Engineering, National Cheng Kung University, Tainan 701, Taiwan, R.O.C. (e-mail: super@mail.ncku.edu.tw).}}

\maketitle

\begin{abstract}
    This paper presents a coding approach for achieving omnidirectional transmission of certain common signals in massive multi-input multi-output (MIMO) networks such that the received power at any direction in a cell remains constant for any given distance. Specifically, two-dimensional (2D) Golay complementary array set (GCAS) can be used to design optimal massive MIMO precoding matrix so as to achieve omnidirectional transmission due to its complementary autocorrelation property. In this paper, novel constructions of new 2D GCASs with arbitrary array lengths are proposed. Our key idea is to carefully truncate the columns of certain larger arrays generated by 2D generalized Boolean functions. Finally, the power radiation patterns and numerical results are provided to verify the omnidirectional property of the GCAS-based precoding. The error performances of the proposed precoding scheme are presented to validate its superiority over the existing alternatives.
\end{abstract}

\begin{IEEEkeywords}
Generalized Boolean function (GBF), Golay complementary array pair (GCAP), Golay complementary array set (GCAS), omnidirectional precoding (OP), uniform rectangular array (URA).
\end{IEEEkeywords}
\section{Introduction}
Complementary pairs/sets of sequences have attracted a sustained research interest owing to their zero aperiodic correlation sums properties. To be specific, a Golay complementary pair (GCP) refers to a pair of equal-length sequences whose summation of aperiodic autocorrelations is zero except at the zero time-shift \cite{Golay}. Such a concept was extended to Golay complementary set (GCS) with constituent sequences of more than 2 by Tseng and Liu in \cite{Golay_sets}. Furthermore, a maximum collection of GCSs is called a set of complete complementary code (CCC) \cite{N_shift} if any two different GCSs have zero aperiodic cross-correlation sums for all time-shifts. In the literature, GCSs and CCCs have been widely used for radar sensing \cite{pezeshki2008doppler}, channel estimation\cite{CS_CE}, precoding for massive multi-input multi-output (MIMO) \cite{su2019omnidirectional}, peak-to-average power ratio (PAPR) reduction in orthogonal frequency division multiplexing (OFDM) \cite{Liu_QAM, Nee2,Golay_RM, Paterson_00,schmidt, ChenAAECC06, chen2016complementary}, interference-free multicarrier code division multiple access \cite{Liu_TWC, Chen_CDMA,Sue_CDMA,Bell_CDMA}, and many other applications \cite{CS_sync, Golay_power1}.

Recently, there is a surge of research attention to study two-dimensional (2D) Golay complementary array sets (GCASs) [18]-[23], each having zero aperiodic autocorrelation sums property for two directions of shifts (compared to conventional GCSs and CCCs with time-shifts only). An important application of the 2D GCASs is for omnidirectional transmission in MIMO communication systems with a uniform rectangular array (URA) configuration \cite{lu2020omnidirectional}, \cite{li2021construction}. In massive MIMO systems, some common messages (e.g., reference signals, synchronization signals, control signals, etc.) need to be power-uniformly broadcasted to all the angles within the whole cell. In this paper, we consider space-time block code (STBC) for the harvesting of the diversity gain. At the base station (BS), the STBC encoded symbols are assigned to several streams and then mapped onto the antenna arrays in URA by certain 2D GCASs assisted precoding matrices to achieve uniform power radiation at any angle.

On the other hand, since a large number of antennas are considered in massive MIMO systems, a huge pilot overhead may be needed to acquire the channel state information (CSI). As pointed out in \cite{meng2015omnidirectional}, this can be alleviated by omnidirectional precoding (OP) based transmission. For uniform linear arrays (ULAs), Zadoff-Chu (ZC) sequences were adopted to satisfy the requirements of the omnidirectional property. However, \cite{meng2015omnidirectional}  only considered the omnidirectional transmission in certain directions. Later in \cite{su2019omnidirectional}, GCSs and CCCs based OP matrices were proposed to meet the requirement of omnidirectional transmission across all directions.

In \cite{jiang2019autocorrelation, lu2020omnidirectional, li2021construction, matsufuji2004construction}, 2D GCASs were employed for precoding matrices in URAs by applying interleaving and Kronecker-product to existing 1D sequences or 2D arrays. As a result, the array sizes of 2D GCASs are only feasible for certain lengths. A construction of 2D GCASs of array size $p^n\times p^m$ was proposed in \cite{wang2020constructions} by using permutation ploynomials (PPs) functions and 2-level autocorrelation sequences, where $p$ is a prime number, $m,n$ are two positive integers, and $p,m,n>0$. Furthermore, a unifying construction framework for 2D GCASs was developed in \cite{wang2021new} by a multivariate polynomial matrix from certain seed para-unitary (PU) matrices. In \cite{pai2020constructions,pai_2022TCOM_2D}, Pai and Chen proposed direct constructions of 2D Golay complementary array pairs (GCAPs) and GCASs with array size $2^n\times 2^m$ from 2D generliazed Boolean functions (GBFs) \cite{Liu_2014} where $n,m$ are integers and $n,m\geq 2$. 2D GCAP can be regarded as a case of 2D GCAS when the set size is equal to 2.  Moreover, Pai {\it et al.} \cite{pai2022constructions} proposed a direct construction of 2D CCCs with array size $2^n\times 2^m$, which have ideal autocorrelations and cross-correlations. Later, Liu $et~al.$ \cite{liu2022constructions} proposed a construction of GCASs with array size $p^n\times p^m$ by using 2D multivariable functions, where $p$ is a prime number, $n,m$ are integers, and $n,m\geq 2$. Based on \cite{pai2020constructions}, \cite{shen2022three} developed a direct construction of GCASs with set size $4$ and array size $2^n\times (2^{m-1}+2^v)$ by using 2D GBFs, where $n,m,v$ are positive number with $n,m\geq 2, \text{and } 0\leq v\leq m-1$.

The aforementioned research efforts are generally driven by the need of highly flexible array sizes of 2D GCASs. Motivated by this, we aim for generating new GCASs with arbitrary array lengths. The key idea of our proposed constructions is to carefully truncate some columns of the certain larger arrays generated by 2D GBFs. Thus, our proposed GCASs can be applied to URAs with various array sizes. In addition, the proposed GCASs can be directly generated from 2D GBFs without the requirements of any specific sequences or tedious sequence operations.
In Table \ref{table3}, we compare the existing parameters of 2D GCASs with our proposed ones.


The remainder of this paper is defined as follows. Section \ref{sec:def} discusses notations, definitions, system models, and the omnidirectional transmission in MIMO systems. Section \ref{sec:constructions of GCASs} describes our proposed constructions of 2D GCASs. Section \ref{sec:simulation} shows the power radiation pattern and bit error rate (BER) performance based on our proposed 2D GCASs precoding. Finally, Section \ref{sec:conclusion} presents the conclusion.
\begin{table*}[ht]
    \centering
    \caption{A COMPARISON OF Constructions FOR 2D GCASs} \label{table3}
    \begin{tabular}{|l|l|l|}
        \hline
        \multicolumn{1}{|c|}{Construction}                                                     & \multicolumn{1}{c|}{Parameters}                                                                                                                                                        & \multicolumn{1}{c|}{Approaches}                                                                              \\ \hline
        \multicolumn{1}{|c|}{\cite[Th. 5]{wang2021new}}                                                             & $(N,N^{n},N^{m}),N,n,m>0$                                                                                                                                                              & {\multirow{2}{*}{\qquad Seed PU matrices}}                                                                 \\\cline{1-2}
        \multicolumn{1}{|c|}{\cite[Th. 7]{wang2021new}}                                                              & $(2^k,2^{kn},2^{km}),n,m,k>0$                                                                                                                                                          &                                                                                                            \\ \hline
        \multicolumn{1}{|c|}{\cite[Th. 4]{wang2020constructions}}                                                    & $(p,p^{n},p^{m})$, prime $p,n,m>0$                                                                                                                                                     & {\multirow{2}{*}{\begin{tabular}[c]{@{}c@{}c@{}}PPs and 2-level\\ autocorrelation sequences\end{tabular}}} \\ \cline{1-2}
        \multicolumn{1}{|c|}{\cite[Th. 6]{wang2020constructions}}                                                    & $(p^k,p^{kn},p^{km}),$ prime $p,k,n,m>0$                                                                                                                                               &                                                                                                            \\ \hline
        \multicolumn{1}{|c|}{\cite[Th. 1]{liu2022constructions}}                                                     & $(p_1^{k_1}p_2^{k_2},p_1^{n},p_2^{m}),$ primes $p_1,p_2$                                                                                                                               & {\multirow{2}{*}{2D multivariable functions}}                                                             \\ \cline{1-2}
        \multicolumn{1}{|c|}{\cite[Th. 2]{liu2022constructions}}                                                     & $(p^k,p^{n},p^{m})$, prime $p$, $n+m\geq k>0$                                                                                                                                          &                                                                                                            \\ \hline
        \multicolumn{1}{|c|}{\cite{pai2020constructions,pai_2022TCOM_2D,pai2022constructions}} & $(2^k,2^{n},2^{m})$, $n,m\geq k>0, \text{and}~k>0$                                                                                                                                     & {\multirow{5}{*}{\qquad\quad 2D GBFs}}                                                                    \\ \cline{1-2}
        \multicolumn{1}{|c|}{\cite{shen2022three}}                                             & $(4,2^{n},2^{m-1}+2^{v})$, $n,m\geq2, \text{and}~k>0$                                                                                                                                &                                                                                                            \\ \cline{1-2}
        \multicolumn{1}{|c|}{Th. \ref{thm:GCAS}}                                               & {\begin{tabular}[c]{@{}c@{}c@{}}
        $(2^{k+1},2^n,2^{m-1}+\sum_{\alpha=1}^{k-1}d_{\alpha}2^{m-k+\alpha-1}+d_02^v),$\\ $k<m,~0\leq v\leq m-k,~d_{\alpha}\in\{0,1\}$\end{tabular}} &                                                                                                            \\ \cline{1-2}
        \multicolumn{1}{|c|}{Th. \ref{thm:GCAS_2}}                                             & {\begin{tabular}[c]{@{}c@{}c@{}}
        $(2^{k+1},2^n,2^{m-1}+\sum_{\alpha=1}^{k-1}d_{\alpha}2^{\pi_{1}(m-k+\alpha)-1}+d_02^v),$\\ $k<m,~0\leq v\leq m-k,~d_{\alpha}\in\{0,1\}$\end{tabular}}                                                                                                                                                                                        &                                                                                                            \\ \hline
    \end{tabular}
\end{table*}
\section{Preliminaries and Definitions}\label{sec:def}
\subsection{Notations}
Throughout this paper, we present the notations in the following:
\begin{itemize}
    \item $({\bm a})_i$ refers to the $i$-th element of the vector $\bm a$.
    \item $({\bm A})_{i,j}$ denotes the $(i,j)$-th element of the array $\bm A$.
    \item $(\cdot)^H$ refers to the conjugate transpose.
    \item diag($\bm A$) refers to the column vector composed of the main diagonal of $\bm A$.
    \item $(\cdot)^*$ refers to the complex conjugation of an element.
    \item $(\cdot)^T$ refers to the transpose.
    \item $\text{vec}(\cdot)$ express stacking one column of the matrix into one another column.
    \item $\bm 1$ is a vector whose elements are all 1.
    \item Let $\xi =e^{2\pi\sqrt{-1}/q}$.
    \item In this paper, $q$ is an even number.
\end{itemize}
Let ${\bm X}$ and ${\bm Y}$ be two arrays of size $L_1\times L_2$. Then ${\bm X}$ and ${\bm Y}$ can be stated as
\begin{equation}
\begin{aligned}
 {\bm X}=(X_{g,i}),~{\bm Y}=(Y_{g,i}),
\end{aligned}
  \label{eq:arrayC}
\end{equation}
where $g=0,1,\cdots,L_1-1$ and $i=0,1,\cdots,L_2-1$.
\begin{defin}
Given two arrays ${\bm X}$ and ${\bm Y}$ of size $L_1\times L_2$, the {\em 2D aperiodic cross-correlation function} (AACF) is defined by
\end{defin}
\begin{align}
    \rho \left( {{\mathbf{X}},{\mathbf{Y}};{u_1},{u_2}} \right) =
 \begin{cases} \sum\limits_{g = 0}^{{L_1} - 1 - {u_1}} {\sum\limits_{i = 0}^{{L_2} - 1 - {u_2}} {{Y_{g + {u_1},i + {u_2}}} {X^{*}_{g,i}}} },0 \leq {u_1} < {L_1}, \\ 0 \leq {u_2} < {L_2}; \\ \sum\limits_{g = 0}^{{L_1} - 1 - {u_1}} {\sum\limits_{i = 0}^{{L_2} - 1 - {u_2}} {{Y_{g + {u_1},i}} {X^{*}_{g,i - {u_2}}}} },0 < {u_1} < {L_1}, \\ - {L_2} < {u_2} < 0; \\ \sum\limits_{g = 0}^{{L_1} - 1 - {u_1}} {\sum\limits_{i = 0}^{{L_2} - 1 - {u_2}} {{Y_{g,i}}  {X^{*}_{g - u_1,i - {u_2}}}} }, - {L_1} < {u_1} < 0, \\ - {L_2} < {u_2} < 0; \\ \sum\limits_{g = 0}^{{L_1} - 1 + {u_1}} {\sum\limits_{i = 0}^{{L_2} - 1 - {u_2}} {{Y_{g,i + {u_2}}} {X^{*}_{g - {u_1},i}}} }, - {L_1} < {u_1} < 0, \\ 0 < {u_2} < {L_2}. \end{cases}
\end{align}
When $\bm X= \bm Y$, then it is called {\em 2D aperiodic autocorrelation function} (AACF) and denoted by $\rho(\bm X; u_1,u_2)$. If taking $L_1=1$, two 2D arrays $\bm X$ and $\bm Y$ are degraded as a 1-D sequence ${\bm X}=X_i$ for $i=0,1,\cdots,L_2-1$ and ${\bm Y}=Y_i$ for $i=0,1,\cdots,L_2-1$, respectively. Then the 1-D AACF of 1-D sequence ${\bm X}$ is related by
\begin{equation}
\begin{aligned}
\rho ({\boldsymbol X};u)=\begin{cases}\sum\limits_{i=0}^{L_{2}-1-u}{X_{i+u}}{X_{i}^{*}},\quad 0 \leq u\leq L_2-1;\\
\sum\limits_{i=0}^{L_{2}-1+u}{X_{i}}{X_{i-u}^{*}},\quad -L_2+1 \leq u<0.\end{cases}
\end{aligned}
\end{equation}
In this paper, $q$-PSK modulation is employed. Thus, $\bm x$ and $\bm y$ denote $q$-ary arrays and (\ref{eq:arrayC}) is expressed as
\begin{equation}
\begin{aligned}
&{\boldsymbol X}=(X_{g,i}) =(\xi ^{x_{g,i}})=\xi ^{\boldsymbol x};\\
&{\boldsymbol Y}=(Y_{g,i}) =(\xi ^{y_{g,i}})=\xi ^{\boldsymbol y},
\label{eg:define_c}
\end{aligned}
\end{equation}
where ${\boldsymbol x} = (x_{g,i}),{\boldsymbol y} = (y_{g,i})$, and $x_{g,i},~y_{g,i}\in \mathbb{Z}_q=\{0,1,\cdots,q-1\}$ for $ 0 \leq g < {L_1},~0 \leq i < {L_2}$.
Consider a set of $N$ $L$-length sequences can be represented as
\begin{equation*}
\begin{aligned}
{ C} = \{{\bm X}_0,{\bm X}_1,\cdots,{\bm X}_{N-1}\}\\
\end{aligned}
\end{equation*}
where
\begin{equation*}
\begin{aligned}
{\boldsymbol X}_n = \left({ X}_{n,0},{ X}_{n,1},\cdots,{ X}_{n,L-1}\right)\\
\end{aligned}
\end{equation*}
for $n=0,1,\cdots,N-1$.
\begin{defin}
\cite{Golay_power1} If a set $C$ consisting of $N$ sequences of length $L$ satisfies
\begin{equation}
\begin{aligned}
  \sum_{k=0}^{N-1}\rho({\bm X_k};u)=\begin{cases}
NL,\quad u=0;\\
0,\qquad u\neq 0,\end{cases}
\end{aligned}
\end{equation}
\end{defin}
then the set $C$ is called a {\em Golay complementary set} of size $N$, denoted by $(N,L)$-GCS. The GCP can be regarded as a special case of the GCS by setting $N=2$.
\begin{defin}
For a GCP $(\bm X_0, \bm X_1)$, if another GCP $(\bm Y_0, \bm Y_1)$ meets the following condition:
\begin{equation}
\begin{aligned}
\rho(\bm X_0, \bm Y_0; u)+\rho(\bm X_1, \bm Y_1; u)=0 ,~\text{for all}~ u,
\end{aligned}
\label{eq:mate}
\end{equation}
then the two GCPs are called the {\em Golay complementary mate} of each other.
\end{defin}
\begin{defin}
A pair of arrays $\bm X$ and $\bm Y$ of array size $L_1\times L_2$ is called a 2D {\em Golay complementary array pair} if
\begin{equation}
\begin{aligned}
&\rho(\bm X;u_1,u_2)+\rho(\bm Y;u_1,u_2)
=\begin{cases}
2L_1L_2,\quad u_1=u_2=0;\\
0,~\quad\qquad u_1\neq0~\text{or}~u_2\neq0.\end{cases}
\end{aligned}
\label{eq:GCAP}
\end{equation}
\end{defin}
\begin{defin}
 Let the array set $G=\{{\bm X}_0,{\bm X}_1,\cdots,{\bm X}_{N-1}\}$ where each array in set $G$ is of size $L_1\times L_2$. If the array set $G$ satisfies
\begin{equation}
\begin{aligned}
\sum_{k=0}^{N-1}\rho({\bm X}_k;u_1,u_2)=\begin{cases}
NL_1L_2, \quad \quad u_1=u_2=0;\\
0,~\qquad\qquad u_1\neq0~\text{or}~u_2\neq0,
\end{cases}
\end{aligned}
  \label{eq:GCAS}
\end{equation}
\end{defin}
the set $G$ is called the {\em Golay complementary array set} of set size $N$ denoted by  $(N,L_1,L_2)$-GCAS where $L_2$ is defined as the length of the GCAS. If $N=2$, the GCAS $G$ is degraded as a GCAP.
\subsection{Generalized Boolean Functions}
 A 2D generalized Boolean function (GBF) $f$ in $n+m$ binary variables  $y_1,y_2,\cdots,y_n,\\x_1,x_2,\cdots,x_m, $ is a function mapping: $\mathbb{Z}_2^{n}\times \mathbb{Z}_2^{m} \rightarrow \mathbb{Z}_q$, where $x_i,y_g\in\{0,1\}$ for $i=1,2,\cdots,m$ and $g=1,2,\cdots,n$. A monomial of degree $r$ is given by any product of $r$ distinct variables among  $y_1,y_2,\cdots,y_n,x_1,x_2,\cdots,x_m$. For instance, $x_1x_3y_1y_2$ is a monomial of degree $4$. Next, the variables $z_1,z_2,\cdots,z_{n+m}$ are defined as
\begin{equation}\label{Z-define}
\begin{aligned}
     z_l=\begin{cases}
          y_l\quad \text{if}~1\leq l \leq n;\\
           x_{l-n} \quad \text{if}~n< l \leq m+n,
          \end{cases}
\end{aligned}
\end{equation}
which are useful for our proposed constructions. For a 2D GBF with $n+m$ variables, the 2D $\mathbb{Z}_q$-valued array
\begin{equation}
{\boldsymbol f}= \begin{pmatrix}f_{0,0}& f_{0,1} & \cdots & f_{0,2^{m}-1}\\ f_{1,0} & f_{1,1} & \cdots & f_{1,2^{m}-1}\\ \vdots & \vdots & \ddots & \vdots \\ f_{2^{n}-1,0} & f_{2^{n}-1,1} & \cdots & f_{2^{n}-1,2^{m}-1} \end{pmatrix}
\end{equation}
of size $2^n\times2^m$ is given by letting $f_{g,i}=f((g_1,g_2,\cdots,g_n),(i_1,i_2,\cdots,i_m))$, where $(g_1,g_2,\cdots,g_n)$ and $(i_1,i_2,\cdots,i_m)$ are binary vector representations of integers $g=\sum_{h=1}^n g_{h}2^{h-1}$ and $i=\sum_{j=1}^n i_{j}2^{j-1}$, respectively.
\begin{eg}\label{eg:2D_GBF}
Taking $q=4,~n=2,$ and $m=3$ for example, the 2D GBF is given as $f=3z_5z_4+z_2z_3+2z_2$. Then the array $\bm f$ of size $4\times 8$ corresponding to $f$ can be obtained, i.e.,
\begin{equation}
{\boldsymbol f}=
\begin{pmatrix}
0& 0 & 0 & 0 & 0& 0 & 3 & 3\\
0& 0 & 0 & 2 & 1& 1 & 3 & 3\\
2& 3 & 2 & 3 & 2& 3 & 1 & 2\\
2& 3 & 2 & 3 & 2& 3 & 1 & 2
\end{pmatrix}.
\end{equation}
\end{eg}
The GBF $f$ can be rewritten as $f=3x_3x_2+y_2x_1+2y_2$. In this paper, we consider the array size $\neq 2^n\times 2^m$. Hence, we define the truncated array ${\bm f}^{(L)}$ corresponding to the 2D GBF $f$ by ignoring the last $2^m-L$ columns of the corresponding array $\bm f$.
\begin{eg}
Following the same notations given in Example \ref{eg:2D_GBF}, the truncated array ${\bm f}^{(6)}$ is given by
\begin{equation}
{\boldsymbol f}^{(6)}=
\begin{pmatrix}
0& 0 & 0 & 0 & 0& 0 \\
0& 0 & 0 & 2 & 1& 1 \\
2& 3 & 2 & 3 & 2& 3 \\
2& 3 & 2 & 3 & 2& 3
\end{pmatrix}.
\end{equation}
\end{eg}
For simplicity, we use $\bm f $ to stand for ${\bm f}^{(L)}$ when $L$ is known.
\subsection{System Model}
\begin{figure*}[!t]
\centering
\begin{center}
\extrarowheight=3pt
\includegraphics[width=150mm]{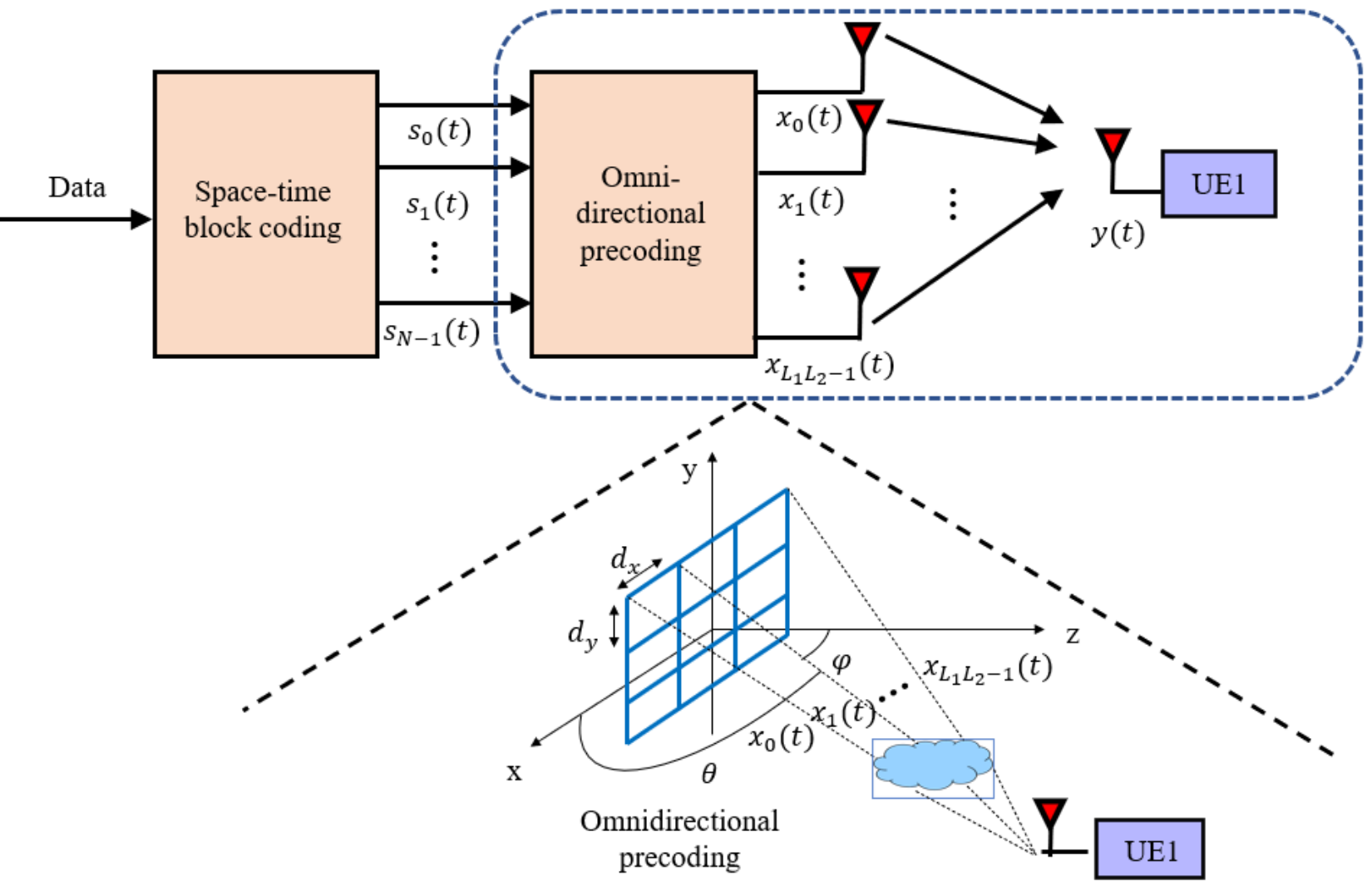}
\caption{Diagram of data transmission through STBC encoding and omnidirectional precoding.}
\label{P1}
\end{center}
\end{figure*}

Considering downlink transmission from a BS to UEs where each has one single antenna, we suppose that the number of antennas at the BS is $M=L_1\times L_2$, i.e., the URA consists of $L_1$ rows and $L_2$ columns. Fig. \ref{P1} illustrates the diagram of data downlink transmission. For an $L_1\times L_2$ URA, the steering matrix ${\bm A}(\varphi,\theta)$ at the direction $(\varphi,\theta)$ with the $(g,i)$-th entry can be expressed as
\begin{equation}
\begin{aligned} {({\bm A}(\varphi,\theta)})_{g,i} =& e^{-j\frac{2\pi }{\lambda }g d_y\sin \varphi \sin \theta -j\frac{2\pi }{\lambda }i d_x\sin \varphi \cos \theta }, \\ &\text{for}\;\;g = 0,1,\ldots,L_1-1,\;\; i = 0,1,\ldots,L_2-1,\\ &{\kern20.0pt}\theta \in [0,2\pi ],\,\, \varphi \in [0,\pi /2],
\label{steeringvector}
\end{aligned}
\end{equation}
where $d_x$ and $d_y$ denote the vertical antenna and horizontal antenna inter-element spacings of the URA, respectively, and $\lambda$ denotes the carrier wavelength.
To enhance the spatial diversity and communication reliability, the STBC signal transmission scheme is used. The $N\times M$ STBC is given by
\begin{equation}
\begin{aligned} \boldsymbol{S}\triangleq \left(\begin{array}{llll}s_0(0) & s_0(1) & \cdots \ &s_0(M-1)\\ s_1(0) & s_1(1) & \cdots \ &s_1(M-1)\\ \vdots & \vdots & \ddots & \vdots \\ s_{N-1}(0) & s_{N-1}(1) & \cdots \ &s_{N-1}(M-1) \end{array}\right) \in {\mathbb {C}}^{N\times M}
\end{aligned}
\end{equation}
where ${\mathbb {C}}^{N\times M}$ refers to the $N$-by-$M$ complex space and $s_n(t)$ denotes the $(n,t)$-th element of the STBC at time instant $t$ for $t=0,1,\cdots,M-1$. We define the precoding matrix ${\bm W}_n$ of size $L_1\times L_2$. The encoded symbols is given by
\begin{equation}
\begin{aligned}
{\bm x}(t)=(x_0(t),x_1(t),\cdots,x_{L_1L_2-1}(t))^{{T}}=\text{vec}\left(\sum_{n=0}^{N-1}{\bm W}_n\cdot s_n(t)\right),~ \text{for }t=0,1,\cdots,M-1,
\label{trans_signal}
\end{aligned}
\end{equation}
which are transmitted by the $L_1L_2$ antennas of the URA. In the light-of-sight (LOS) channel without multipaths, the received signal at the direction $(\varphi,\theta)$ can be written as
\begin{equation}
\begin{aligned}
y(t)=\sum _{n=0}^{N-1} \left({\text{vec}}({\bm A}(\varphi,\theta))^T {\text{vec}}({\bm W}_n)\right) \cdot s_n(t) +\eta(t),\, t=0,{\ldots },M-1,
\label{received_singal}
\end{aligned}
\end{equation}
where $\eta(t)$ is the additive Gaussian white noise (AWGN) at time instant $t$.

\subsection{Omnidirectional Precoding Matrices Based on 2D Arrays}
In this subsection, we list two necessary requirements for the design of OP matrices. Then, we will connect these two requirements with the conditions of 2D arrays.

{\it Requirement 1 (R1): Omnidirectional transmission.}

We consider the MIMO system with URA. Following (\ref{received_singal}), the received power $E$ at the angle $(\varphi,\theta)$ is represented as
\begin{equation}
\begin{aligned}
E=\sum _{n=0}^{N-1} \left|[{\text{vec}}({\bm A}(\varphi,\theta))^T {\text{vec}}({\bm W}_n)]\right|^2.
\label{power radiation}
\end{aligned}
\end{equation}
Therefore, to satisfy the omnidirectional transmission in the whole cell, (\ref{power radiation}) must be constant for all $\varphi\text{ and}~\theta$.

{\it Requirement 2 (R2): Equal average power on each antenna.}

To enhance the efficiency of the power amplifier, the average transmission power on all $L_1\times L_2$ antennas is required  to be equal. We define
\begin{equation}
\begin{aligned}
&{\bm W}=\left(\text{vec}({\bm W}_0), \text{vec}({\bm W}_1),\cdots, \text{vec}({\bm W}_{N-1})\right),
\label{vec_GCAS}
\end{aligned}
\end{equation}
where the array size of $\bm W$ is $L_1L_2\times N$. Hence, (\ref{trans_signal}) can be rewritten as
\begin{equation}
\begin{aligned}
{\bm X}=\left({\bm x}(0),{\bm x}(1),\cdots,{\bm x}(M-1)\right)={\bm W}{\bm S}.
\end{aligned}
\end{equation}
Let ${\bm s}(t)$ be the $t$-th column of $\bm S$. Throughout this paper, we assume $\mathbb{E}\left[{\bm s}{(t)}{\bm s}{(t)}^H\right]$=${\bm I}_N$. The transmitted signal on the $(l_1,l_2)$-th antenna is $(\bm W \bm s)_{l_2L_1+l_1}$. The average power on the $(l_1,l_2)$-th antenna can be expressed as
\begin{equation}
\begin{aligned}
\mathbb{E}\left[|({\bm W}{\bm s})_{l_2L_1+l_1}|^2\right]&=\left({\bm W}\mathbb{E}\left[{\bm s}{(t)}{\bm s}{(t)}^H\right]{\bm W}^H\right)_{l_2L_1+l_1,l_2L_1+l_1}\\
&=({\bm W}{\bm W}^H)_{l_2L_1+l_1,l_2L_1+l_1}.
\label{power on each antenna}
\end{aligned}
\end{equation}
Therefore, the condition to guarantee equal power on each antenna is equivalent to
\begin{equation}
\begin{aligned}
\text{diag}({\bm W}{\bm W}^H)=N{\bm 1}.
\label{power diagonal}
\end{aligned}
\end{equation}
Next, we will derive two sufficient conditions on the precoding matrices to fulfill requirements R1 and R2.
\begin{lemma}
\cite{li2021construction} For an $L_1\times L_2$ URA, if the precoding matrices ${\bm W}_0,{\bm W}_1,\cdots,{\bm W}_{N-1}$ of size $L_1\times L_2$ form an $(N,L_1,L_2)$-GCAS, then the omnidirectional transmission is achieved.
\label{lemma omnidirectional}
\end{lemma}
\begin{lemma}
For an $L_1\times L_2$ URA, if the precoding matrices ${\bm W}_0,{\bm W}_1,\cdots,{\bm W}_{N-1}$ of size $L_1\times L_2$ are unimodular, then the average power on each antenna is equal.
\label{lemma power on each antenna}
\end{lemma}
\begin{IEEEproof}
In order to meet the requirement for equal average power on each antenna, the precoding matrix $\bm W$ must satisfy (\ref{power diagonal}). We let ${\bm w}_i=\text{vec}({\bm W}_i),~\text{for }i=0,1,\cdots,N-1$. Then,
\begin{equation}
\begin{aligned}
{{\text{diag}}}\left({\bm{W}}{{{\bm{W}}}^H}\right)&=\left(\sum_{i=0}^{N-1}\left|({\bm w}_i)_0\right|^2,\ \sum_{i=0}^{N-1}\left|({\bm w}_i)_1\right|^2,\ \cdots,\ \sum_{i=0}^{N-1}\left|({\bm w}_i)_{L_1L_2-1}\right|^2\right)^T\\
&=N{{{\bm{1}}}}
\end{aligned}
\end{equation}
since we have
\begin{equation}
\begin{aligned}
\left|({\bm w}_i)_n\right|^2=1,\qquad \text{for}~i=0,1,\cdots,N-1~\text{and } n=0,1,\cdots,L_1L_2-1.
\label{power each antenna equivalent}
\end{aligned}
\end{equation}
According to (\ref{power diagonal}), the requirement (R2) is fulfilled.
\end{IEEEproof}

In the sequel, the design of OP matrices ${\bm W}_0,{\bm W}_1,\cdots,{\bm W}_{N-1}$ are based on Lemma \ref{lemma omnidirectional} and Lemma \ref{lemma power on each antenna}. That is, our goal is to construct unimodular GCASs with flexible sizes.

\section{GCASs With Flexible Array Size}\label{sec:constructions of GCASs}
In this section, two constructions of 2D GCASs with arbitrary array lengths based on 2D GBFs will be proposed. By recalling the function mapping in (\ref{Z-define}), we present our first theorem in the following.
\begin{theo}\label{thm:GCAS}
For any integers $q$, $m,~n\geq 2$, and $k<m$, $v$ is an integer satisfies $0\leq v\leq m-k$ and let $\pi$ be a permutation of $\{1,2,\cdots m+n-k\}$ satisfying  $\{z_{\pi(1)},z_{\pi(2)},\cdots,z_{\pi(v+n)}\}=\{z_1,z_2,\cdots,z_{v+n}\}$. The 2D generalized Boolean function can be written as
\begin{equation} \label{thm:3}
\begin{aligned}
        f=\frac{q}{2}\left(\sum_{l=1}^{m+n-k-1}z_{\pi (l)}z_{\pi(l+1)}\right)+\sum_{s=1}^{m+n} p_s z_s+p_0
\end{aligned}
\end{equation}
where $p_s\in \mathbb{Z}_q$. The array set
\begin{equation}
\begin{aligned}
        G=\left\{ {\bm f}+\frac{q}{2}\sum_{\alpha=1}^k \lambda_{\alpha}{\bm z}_{ m+n-k+\alpha}+\frac{q}{2}\lambda_{k+1}{\bm z}_{\pi (1)}:\lambda_{\alpha}\in\{0,1\}\right\}
\end{aligned}
\end{equation}
is a $q$-ary $(2^{k+1},2^n,2^{m-1}+\sum_{\alpha=1}^{k-1}d_{\alpha}2^{ m-k+\alpha-1}+d_{0}2^v)$-GCAS where $d_{\alpha}\in \{0,1\}$.
\end{theo}
\begin{IEEEproof}
Without loss of generality, we consider $L_1=2^n$ and $L_2=2^{m-1}+\sum_{\alpha=1}^{k-1}2^{m-k+\alpha-1}+~2^v$. We need to show that
\begin{equation}
\begin{aligned}
    \sum_{{\bm c}\in G}\sum\limits _{g=0}^{L_1-1-u_1}\sum\limits _{i=0}^{L_2-1-u_2}\left(\xi ^{c_{g+u_1,{i+u_2}}-c_{g,i}} \right)=0
\end{aligned}
\end{equation}
for $0\leq u_1< 2^n,~0\leq u_2< 2^{m-1}+\sum_{\alpha=1}^{k-1}2^{m-k+\alpha-1}+~2^v$ and $(u_1,u_2)\neq (0,0)$. Then we let $h=g+u_1$ and $j=i+u_2$ for any integers $g$ and $i$. We also let $(g_1,g_2,\cdots,g_n)$,$(i_1,i_2,\cdots,i_m)$,\\$(h_1,h_2,\cdots,h_n)$, and $(j_1,j_2,\cdots,j_m)$ be the binary representations of $g,i,h$, and $j$, respectively.  For the ease of presentation, we denote
\begin{equation}
\begin{aligned}
        &a_l=\begin{cases}
                g_l\qquad \text{for}~1\leq l\leq n;\\
                i_{l-n}~~ \text{for}~n< l\leq n+m;
              \end{cases}\\
        &b_l=\begin{cases}
                h_l\qquad \text{for}~1\leq l\leq n;\\
                j_{l-n}~~ \text{for}~n< l\leq n+m;
              \end{cases}
\label{define_ab}
\end{aligned}
\end{equation}
In what follows, we consider four cases to show that the above formula holds.

{\it Case 1:} If $a_{\pi(1)}\neq b_{\pi(1)}$, we can find that ${\bm c}^{\prime}={\bm c}+(q/2){\bm z}_{\pi(1)}$ for any array${\bm c}\in G$ satisfying
\begin{equation}
\begin{aligned}
        c_{h,j}-c_{g,i}-{c}^{\prime}_{h,j}\!\!+\!\!{c}^{\prime}_{g,i}= \frac{q}{2}(a_{\pi (1)}- b_{\pi (1)})\equiv \frac{q}{2} \pmod q.
\end{aligned}
\end{equation}
Therefore, we have
\begin{equation}
\begin{aligned}
\xi^{c_{h,j}-c_{g,i}}+\xi^{c^{\prime}_{h,j}-c^{\prime}_{g,i}}=0.
\end{aligned}
\end{equation}

{\it Case 2:} If $a_{m+n-k+\alpha}\neq b_{m+n-k+\alpha}$, we can find that ${\bm c}^{\prime}={\bm c}+(q/2){\bm z}_{m+n-k+\alpha}$ for any array ${\bm c}\in G$. Similar to Case 1, we have
\begin{equation}
\begin{aligned}
\xi^{c_{h,j}-c_{g,i}}+\xi^{c^{\prime}_{h,j}-c^{\prime}_{g,i}}=0.
\end{aligned}
\end{equation}

{\it Case 3:} If $a_{\pi(1)}=b_{\pi(1)}$ and $a_{m+n-k+\alpha} = b_{m+n-k+\alpha}$ for $\alpha =1,2,\cdots,k$.  Suppose that $\alpha^{\prime}$ is the largest integer satisfying $a_{m+n-k+\alpha^{\prime}} = b_{m+n-k+\alpha^{\prime}}=0$ for $\alpha^{\prime}\leq k$. Then we assume $\beta$ is the smallest integer which satisfies $a_{\pi(\beta)}\neq b_{\pi(\beta)}$. Let $a^{\prime}$ and $b^{\prime}$ be integers distinct from $a$ and $b$, respectively, only in one position $\pi(\beta-1)$. In other words, $a^{\prime}_{\pi(\beta-1)}=1-a_{\pi(\beta-1)}$ and $b^{\prime}_{\pi(\beta-1)}=1-b_{\pi(\beta-1)}$. If $1\leq \pi(\beta-1)\leq n$, by using the above definition, we have
        \begin{equation}
            \begin{aligned}
                 & c_{g^{\prime },i}-c_{g,i}                                                                                                \\
                 & =\frac{q}{2}\left(a_{\pi(\beta-2)}g^{\prime }_{\pi(\beta-1)}-a_{\pi(\beta-2)}g_{\pi (\beta-1)}+g^{\prime }_{\pi (\beta-1)}a_{\pi (\beta)}\right. \\
                 & \left. -g_{\pi (\beta-1)}a_{\pi(\beta)}\right)+p_{\pi (\beta-1)}g^{\prime }_{\pi _2(\beta-1)}-p_{\pi (\beta-1)}g_{\pi(\beta-1)}                  \\
                 & \equiv \frac{q}{2}(a_{\pi (\beta-2)}+a_{\pi (\beta)})+p_{\pi(\beta-1)}(1-2g_{\pi(\beta-1)}) \pmod q.
            \end{aligned}
        \end{equation}
        where $a^{\prime}_{\pi(\beta-1)}=g^{\prime }_{\pi(\beta-1)}$ and $a_{\pi(\beta-1)}=g_{\pi(\beta-1)}$. Since $a_{\pi(\beta-2)}= b_{\pi(\beta-2)}$ and $a_{\pi(\beta-1)}= b_{\pi(\beta-1)}$, we have
        \begin{equation}
            \begin{aligned}
                 & c_{h,j}-c_{g,i}-c_{h^{\prime },j}+c_{g^{\prime },i}                     \\
                 & \equiv \frac{q}{2}(a_{\pi (\beta-2)}-b_{\pi (\beta-2)}+a_{\pi (\beta)}-b_{\pi (\beta)}) \\
                 & \quad +p _{\pi (\beta-1)}(2h_{\pi (\beta-1)}-2g_{\pi (\beta-1)})                    \\
                 & \equiv \frac{q}{2}(a_{\pi (\beta)}-b_{\pi (\beta)})\equiv \frac{q}{2} \pmod q
            \end{aligned}
        \end{equation}
        implying $\xi ^{c_{h,j}-c_{g,i}}/\xi ^{c_{h^{\prime },j}-c_{g^{\prime },i}}=-1$. We can also obtain
        \begin{equation}
            \begin{aligned}
                \xi ^{c_{h,j}-c_{g,i}}+\xi ^{c_{h^{\prime },j}-c_{g^{\prime },i}}=0.
            \end{aligned}
        \end{equation}
        If $n<\pi(\beta-1)\leq n+m$, note that $a^\prime_{\pi(\beta-1)}=i^{\prime}_{\pi(\beta-1)-n}$ and $a_{\pi(\beta-1)}=i_{\pi}(\beta-1)-n$ according to (\ref{define_ab}). Following the similar argument as given above, we can get $\xi ^{c_{h,j}-c_{g,i}}+\xi ^{c_{h,j^{\prime }}-c_{g,i^{\prime }}}=0.$

{\it Case 4:}  If $a_{\pi(1)}=b_{\pi(1)}$ and $a_{m+n-k+\alpha} = b_{m+n-k+\alpha}=1$ for $\alpha =1,2,\cdots,k$. We assume $\beta$ is the smallest integer such that $a_{\pi(\beta)}\neq b_{\pi(\beta)}$. Since $a_s=b_s=0$ for $s=v+n+1,v+n+2,\cdots,m+n-k$, we can obtain $\pi(\beta)\leq v+n$ implying $\pi(\beta-1)\leq v+n$. If $1\leq \pi(\beta-1)\leq n$, by following the similar argument as given above, we have $\xi ^{c_{h,j}-c_{g,i}}+\xi ^{c_{h^{\prime },j}-c_{g^{\prime },i}}=0.$ If $n<\pi(\beta-1)\leq v+n$, we have $\xi ^{c_{h,j}-c_{g,i}}+\xi ^{c_{h,j^{\prime }}-c_{g,i^{\prime }}}=0.$
From Cases 1 to 4, the theorem can be proved.
\end{IEEEproof}

\begin{table*}[ht]
\centering
\caption{The Constructed $(4,4,33)$-GCAS in Example \ref{eg:GCAS1}} \label{table1}
\begin{tabular}{|l|l|l|l|l|l|}
\hline
{$
{\bm c}_0=$\tiny$

\left( {\begin{array}{ccccccccccccccccccccccccccccccccccccccccccccccccccccccccccccccccccccccc}
0&	1&	1&	1&	1&	0&	1&	1&	1&	0&	0&	0&	1&	0&	1&	1&	1&	0&	0&	0&	0&	1&	0&	0&	1&	0&	0&	0&	1&	0&	1&	1&	1\\
1&	0&	0&	0&	0&	1&	0&	0&	0&	1&	1&	1&	0&	1&	0&	0&	0&	1&	1&	1&	1&	0&	1&	1&	0&	1&	1&	1&	0&	1&	0&	0&	1\\
1&	0&	0&	0&	0&	1&	0&	0&	0&	1&	1&	1&	0&	1&	0&	0&	0&	1&	1&	1&	1&	0&	1&	1&	0&	1&	1&	1&	0&	1&	0&	0&	0\\
1&	0&	0&	0&	0&	1&	0&	0&	0&	1&	1&	1&	0&	1&	0&	0&	0&	1&	1&	1&	1&	0&	1&	1&	0&	1&	1&	1&	0&	1&	0&	0&	1\\
\end{array}}
\right)
$}\\ \hline{$
{\bm c}_1=$\tiny$
\left( {\begin{array}{cccccccccccccccccccccccccccccccccccccccccccccccccccccccccccccccccccccccc}
0&	0&	1&	0&	1&	1&	1&	0&	1&	1&	0&	1&	1&	1&	1&	0&	1&	1&	0&	1&	0&	0&	0&	1&	1&	1&	0&	1&	1&	1&	1&	0&	1\\
1&	1&	0&	1&	0&	0&	0&	1&	0&	0&	1&	0&	0&	0&	0&	1&	0&	0&	1&	0&	1&	1&	1&	0&	0&	0&	1&	0&	0&	0&	0&	1&	1\\
1&	1&	0&	1&	0&	0&	0&	1&	0&	0&	1&	0&	0&	0&	0&	1&	0&	0&	1&	0&	1&	1&	1&	0&	0&	0&	1&	0&	0&	0&	0&	1&	0\\
1&	1&	0&	1&	0&	0&	0&	1&	0&	0&	1&	0&	0&	0&	0&	1&	0&	0&	1&	0&	1&	1&	1&	0&	0&	0&	1&	0&	0&	0&	0&	1&	1\\
\end{array}}
\right)
$}\\ \hline
{$
{\bm c}_2=$\tiny$
\left( {\begin{array}{ccccccccccccccccccccccccccccccccccccccccccccccccccccccccccccc}
0&	1&	1&	1&	1&	0&	1&	1&	1&	0&	0&	0&	1&	0&	1&	1&	1&	0&	0&	0&	0&	1&	0&	0&	1&	0&	0&	0&	1&	0&	1&	1&	1\\
1&	0&	0&	0&	0&	1&	0&	0&	0&	1&	1&	1&	0&	1&	0&	0&	0&	1&	1&	1&	1&	0&	1&	1&	0&	1&	1&	1&	0&	1&	0&	0&	1\\
0&	1&	1&	1&	1&	0&	1&	1&	1&	0&	0&	0&	1&	0&	1&	1&	1&	0&	0&	0&	0&	1&	0&	0&	1&	0&	0&	0&	1&	0&	1&	1&	1\\
0&	1&	1&	1&	1&	0&	1&	1&	1&	0&	0&	0&	1&	0&	1&	1&	1&	0&	0&	0&	0&	1&	0&	0&	1&	0&	0&	0&	1&	0&	1&	1&	0\\
\end{array}}
\right)
$}\\ \hline {$
{\bm c}_3=$\tiny$
\left( {\begin{array}{ccccccccccccccccccccccccccccccccccccccccccccccccccccccccccccc}
0&	0&	1&	0&	1&	1&	1&	0&	1&	1&	0&	1&	1&	1&	1&	0&	1&	1&	0&	1&	0&	0&	0&	1&	1&	1&	0&	1&	1&	1&	1&	0&	1\\
1&	1&	0&	1&	0&	0&	0&	1&	0&	0&	1&	0&	0&	0&	0&	1&	0&	0&	1&	0&	1&	1&	1&	0&	0&	0&	1&	0&	0&	0&	0&	1&	1\\
0&	0&	1&	0&	1&	1&	1&	0&	1&	1&	0&	1&	1&	1&	1&	0&	1&	1&	0&	1&	0&	0&	0&	1&	1&	1&	0&	1&	1&	1&	1&	0&	1\\
0&	0&	1&	0&	1&	1&	1&	0&	1&	1&	0&	1&	1&	1&	1&	0&	1&	1&	0&	1&	0&	0&	0&	1&	1&	1&	0&	1&	1&	1&	1&	0&	0\\
\end{array}}
\right)
$}\\ \hline
\end{tabular}
\end{table*}

\begin{figure*}[ht]
\centering
\begin{center}
\extrarowheight=3pt
\includegraphics[width=120mm]{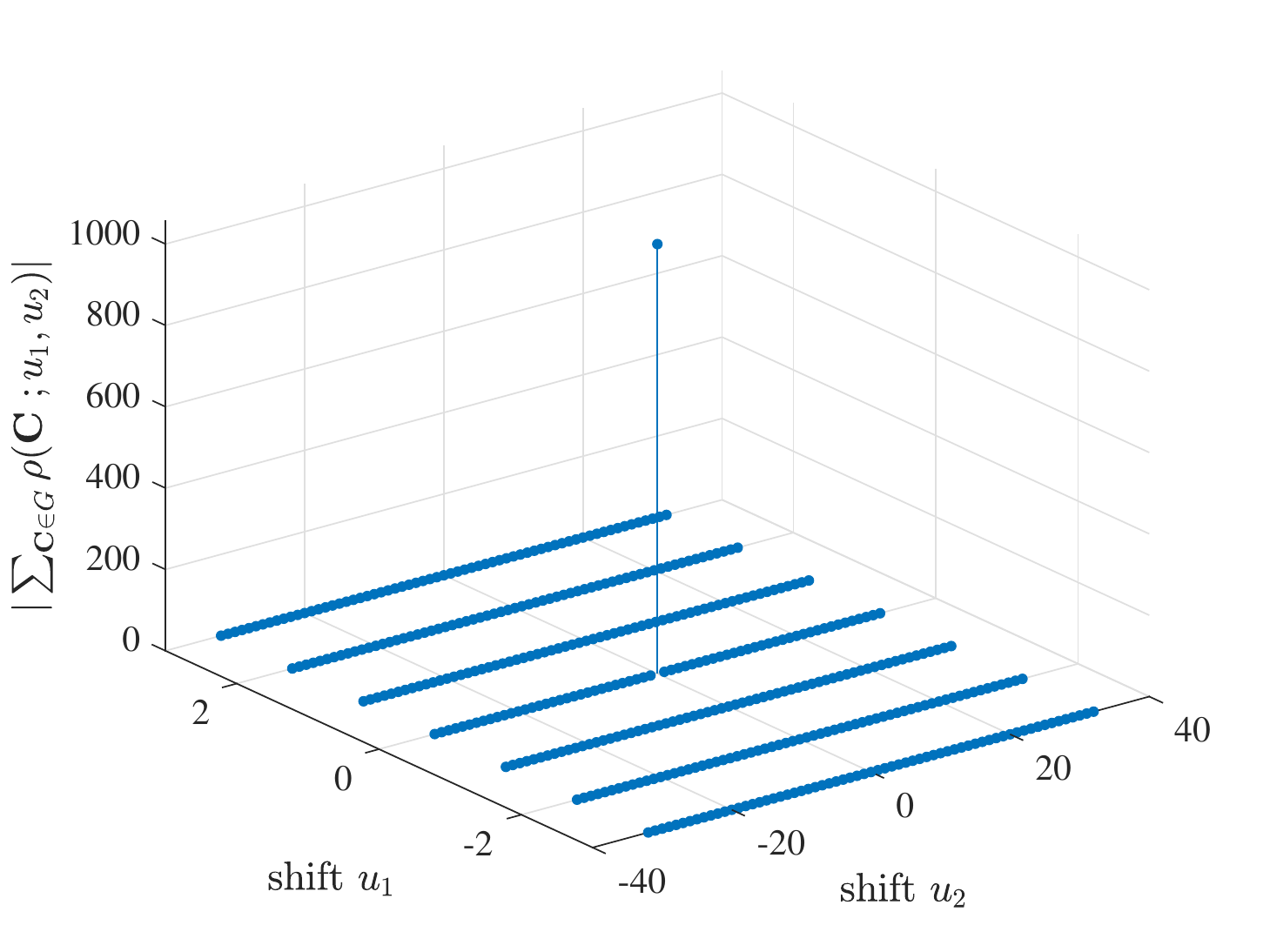}
\caption{The summation of autocorrelations of constituent arrays in the GCAS in Example \ref{eg:GCAS1}.}
\label{GCAS1}
\end{center}
\end{figure*}
\begin{remark}
 The parameter $2^{m-1}+\sum_{\alpha=1}^{k-1}d_{\alpha}2^{ m-k+\alpha-1}+d_{0}2^v$ of the proposed GCASs in Theorem \ref{thm:GCAS} can be any arbitrary length since $m$, $k$, $v$ are flexible and $d_{\alpha}\in \{0,1\}$.
\end{remark}
\begin{eg}
Taking $q=2$, $m=6$, $n=2$, $k=1$, and $v=0$, we let $\pi=(1,2,3,4,5,6,7)$. The generalized Boolean function is $f=z_1z_2 + z_2z_3 +z_3z_4 +z_4z_5 +z_5z_6 +z_6z_7  =x_1x_2+x_2x_3+x_3x_4+x_4x_5+y_1y_2+y_2x_1$ by setting $p_k=0$ for $k=0,1,\ldots, m+n$. The array set $G = \{{\bm f},{\bm f}+{\bm x}_8, {\bm f}+{\bm y}_1, {\bm f}+{\bm x}_8+{\bm y}_1\}$ is a GCAS of size $4$ and the array size is $4\times 33$. We let $G=\{{\bm c}_0,{\bm c}_1,{\bm c}_2,{\bm c}_3\}$ and list the constituent arrays in Table \ref{table1}. Fig. \ref{GCAS1} shows the AACF sum of set $G$ is zero at shift $u_1\neq 0$ or $u_2\neq 0$. Thus, we can find that array set $G$ is a $(4,4,33)$-GCAS.
\label{eg:GCAS1}
\end{eg}

Next, we introduce a lemma which illustrates a construction of $(4,2^n,2^{m-1}+2^v)$-GCAS from 2D GBFs.
\begin{lemma}
\label{lemma_shen2022three}
\cite[Th. 1]{shen2022three} For nonnegative integers $m$, $n$, and $v$ with  $0\leq v<m-1$, let $\pi_1$ be a permutation of $\{1,2,\cdots,m-1\}$ and $\pi_2$ be a permutation of $\{1,2,\cdots,n\}.$ The 2D GBF is given by
\begin{equation}
\begin{aligned}
f=&\frac{q}{2}\left(\sum_{k=1}^{m-2}x_{\pi_1 (k)}x_{\pi_1 (k+1)}+\sum_{k=1}^{n-1}y_{\pi_2 (k)}y_{\pi_2 (k+1)}+x_{\pi_1 (m-1)}x_m+x_m y_{\pi_2 (1)}\right)\\ &+\sum_{l=1}^m p_l x_l+\sum_{s=1}^n\kappa_s y_s+p_0
\label{shen2022three_GBF}
\end{aligned}
\end{equation}
where $p_l,~\kappa_s\in \mathbb{Z}_q$. Then the array set
\begin{equation*}
\begin{aligned}
G=\left\{ {\bm f},{\bm f}+\frac{q}{2}{\bm x}_{\pi_1 (1)}, {\bm f}+\frac{q}{2}{\bm y}_{\pi_2(n)}, {\bm f}+\frac{q}{2}{\bm x}_{\pi_1 (1)}+\frac{q}{2}{\bm y}_{\pi_2(n)} \right\}
\end{aligned}
\end{equation*}
is a $(4,2^n,2^{m-1}+2^v)$-GCAS.
\end{lemma}

Since the set size of the GCAS from Lemma \ref{lemma_shen2022three} is limited to 4, we propose a general construction of 2D GCASs with more flexible array sizes and set sizes which can include Lemma \ref{lemma_shen2022three} as a special case.

\begin{figure*}[!t]
\centering
\begin{center}
\extrarowheight=3pt
\includegraphics[width=120mm]{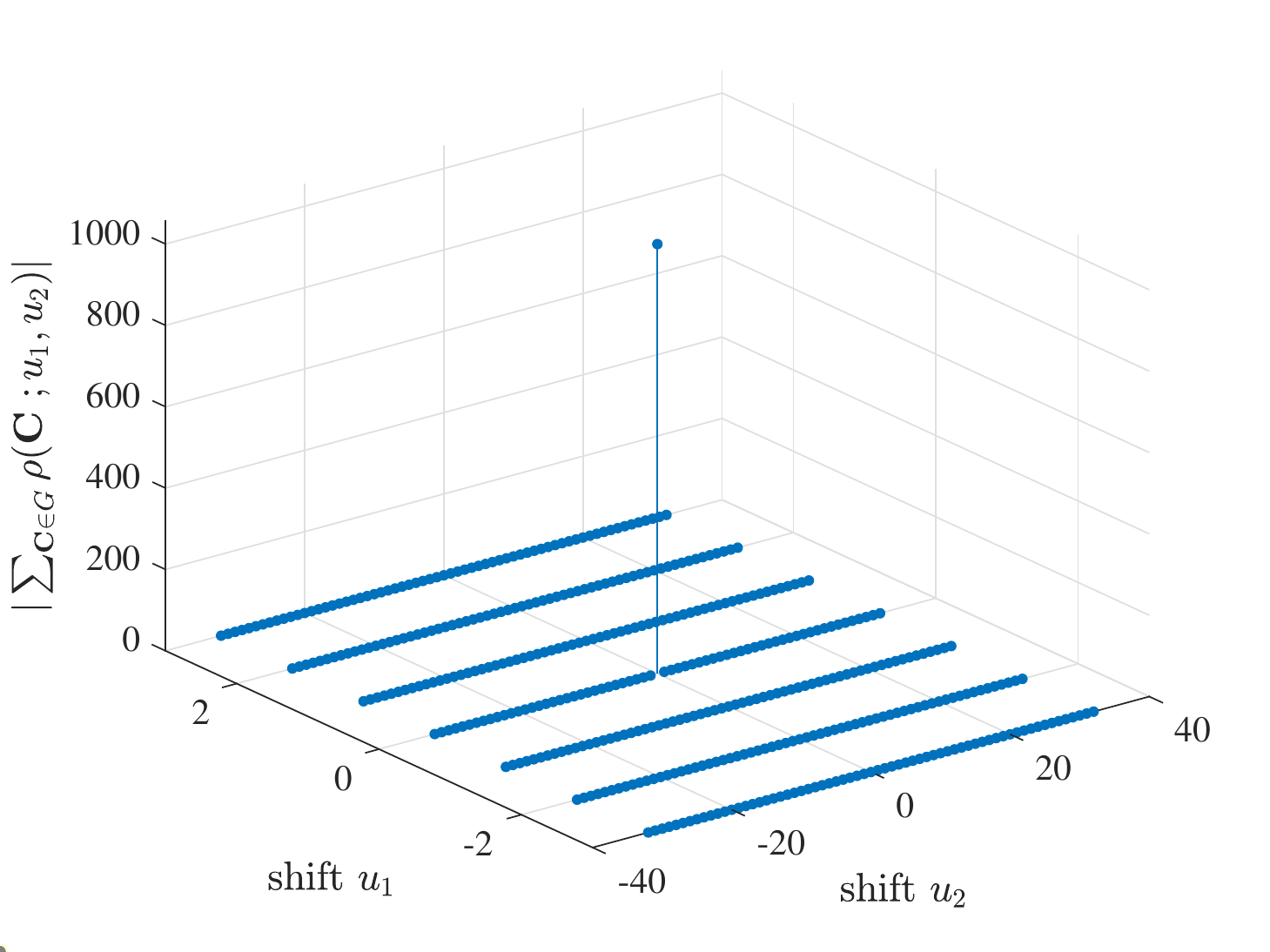}
\caption{The summation of autocorrelations of constituent arrays in the GCAS in Example \ref{eg:GCAS2}.}
\label{fig:GCAS2}
\end{center}
\end{figure*}

\begin{theo}\label{thm:GCAS_2}
For any integers $q$, $m,~n\geq 2$, and $k<m$, $v$ is an integer satisfies $0\leq v\leq m-k$. Assume that $\pi_1$ is a permutation of $\{1,2,\cdots m\}$ and $\pi_2$ is a permutation of $\{1,2,\cdots n\}$. The 2D generalized Boolean function can be written as
\begin{equation}
\begin{aligned}
f=&\frac{q}{2}\left(\sum_{l=1}^{m-k-1}x_{\pi_1 (l)}x_{\pi_1 (l+1)}+\sum_{s=1}^{n-1}y_{\pi_2 (s)}y_{\pi_2 (s+1)}+x_{\pi_1(m)} y_{\pi_2 (n)}\right)\\
&+\sum_{l=1}^{m-k}\mu_l x_{\pi_1(l)}x_{\pi_1(m)}+\sum_{l=1}^m p_l x_k+\sum_{s=1}^n\kappa_s y_s+p_0
\end{aligned}
\end{equation}
where $\mu_l,~p_l,~\kappa_s,\in \mathbb{Z}_q$. The array set
\begin{equation}
\begin{aligned}
G=\left\{ {\bm f}+\frac{q}{2}\sum_{\alpha=1}^{k-1} \lambda_{\alpha}{\bm x}_{\pi_{1}(m-k+\alpha)}+\frac{q}{2}\lambda_{k}{\bm y}_{\pi_2 (1)}+\frac{q}{2}\lambda_{k+1}{\bm x}_{\pi_1 (1)}:\lambda_{\alpha}\in\{0,1\}\right\}
\end{aligned}
\end{equation}
is a $q$-ary $(2^{k+1},2^n,2^{m-1}+\sum_{\alpha=1}^{k-1}d_{\alpha}2^{\pi_{1}(m-k+\alpha)-1}+d_{0}2^v)$-GCAS where $d_{\alpha}\in \{0,1\}$ if the following three conditions hold.

\begin{itemize}
\item[(C1)] $\{\pi_1(1),\pi_1(2),\cdots,\pi_1(v)\}=\{1,2,\cdots,v\}$ if $v>0$;
\item[(C2)] $\pi_{1}(m-k+\alpha)<\pi_{1}(m-k+\alpha+1)$ for $1\leq \alpha\leq k-1$ where $\pi_{1}(m)=m$;
\item[(C3)] For $1\leq \alpha\leq k-1$ and $2\leq \beta\leq m-k$, if $\pi_{1}(\beta)<\pi_{1}(m-k+\alpha)$, then $\pi_{1}(\beta-1)<\pi_{1}(m-k+\alpha)$.
\end{itemize}
\end{theo}
\begin{IEEEproof}
Similarly, we consider $L_1=2^n$ and $L_2=2^{m-1}+\sum_{\alpha=1}^{k-1}2^{\pi_{1}(m-k+\alpha)-1}+~2^v$. Then we would like to prove that
 \begin{equation}
\begin{aligned}
\sum_{{\bm C}}\rho({\bm C};u_1,u_2)=\sum_{{\bm c}\in G}\sum\limits _{g=0}^{L_1-1-u_1}\sum\limits _{i=0}^{L_2-1-u_2}\left(\xi ^{c_{g+u_1,{i+u_2}}-c_{g,i}} \right)=0
\end{aligned}
\end{equation}
for $0\leq u_1< 2^n,~0\leq u_2 < 2^{m-1}+\sum_{\alpha=1}^{k-1}2^{\pi_{1}(m-k+\alpha)-1}+~2^v$ and $(u_1,u_2)\neq (0,0)$. From (\ref{eg:define_c}) we can find that
 \begin{equation}
\begin{aligned}
{\bm c}=\frac{q}{2}&\left(\sum_{l=1}^{m-k-1}{\bm x}_{\pi_1 (l)}{\bm x}_{\pi_1 (l+1)}+\sum_{s=1}^{n-1}{\bm y}_{\pi_2 (s)}{\bm y}_{\pi_2 (s+1)}+{\bm x}_{\pi_1(m)} {\bm y}_{\pi_2 (n)}\right)\\
&+\sum_{l=1}^{m-k}\mu_l {\bm x}_{\pi_1(l)}{\bm x}_{\pi_1(m)}+\sum_{l=1}^m p_l{\bm x}_l+\sum_{s=1}^n\kappa_s {\bm y}_s+p_0\cdot {\bm 1}.
\end{aligned}
\end{equation}
Then we let $h=g+u_1$ and $j=i+u_2$ for any integers $g$ and $i$. Next, we discuss seven cases to complete the proof.

{\it Case 1:} Assuming $u_1> 0$, $u_2\geq 0$, and $g_{\pi_{2}(1)}\neq h_{\pi_{2}(1)}$, we can find an array ${\bm c}^{\prime}={\bm c}+(q/2){\bm y}_{\pi_2(1)}\in G$ for any array ${\bm c}\in G$. Therefore, we can obtain
\begin{equation}
\begin{aligned}
c_{h,j}-c_{g,i}-{c}^{\prime}_{h,j}\!\!+\!\!{c}^{\prime}_{g,i}= \frac{q}{2}(g_{\pi _{2}(1)}- h_{\pi _{2}(1)})\equiv \frac{q}{2} \pmod q
\end{aligned}
\end{equation}
Since $g_{\pi_{2}(1)}\neq h_{\pi_{2}(1)}$, we have
\begin{equation}
\begin{aligned}
\xi ^{c_{h,j}-c_{g,i}}/\xi ^{{c}^{\prime}_{h,j}-{c}^{\prime}_{g,i}}=\xi^{\frac{q}{2}}=-1.
\label{q/2}
\end{aligned}
\end{equation}
Thus,
\begin{equation}
\begin{aligned}
    \xi ^{c_{h,j}-c_{g,i}}+\xi ^{{c}^{\prime}_{h,j}-{c}^{\prime}_{g,i}}=0.
\end{aligned}
\end{equation}

{\it Case 2:} If $u_1>0,~u_2\geq0$, and $g_{\pi_2(1)}= h_{\pi_2(1)}$. Let $\beta$ be the smallest integer such that $g_{\pi_2(\beta)}\neq h_{\pi_2(\beta)}$. We define $g^{\prime}$ and $h^{\prime}$ are two integers which are distinct from $g$ and $h$ only in one position $\pi_2(\beta-1)$, respectively. Then, similar to Case 2 of Theorem \ref{thm:GCAS}, we have
\begin{equation}
\begin{aligned}
    \xi ^{c_{h,j}-c_{g,i}}+\xi ^{c_{h^{\prime},j}-c_{g^{\prime},i}}=0.
\end{aligned}
\end{equation}

{\it Case 3:}  We suppose $i_m\neq j_m$, $u_1=0$ and $u_2>0$. We let $g^{\prime}$ be an integer distinct from $i$ only in one position, i.e., $g^{\prime}_{\pi_2(n)}=1-g_{\pi_2(n)}$. Similar to Case 3 of Theorem \ref{thm:GCAS}, we have $\xi^{c_{g,j}-c_{g,i}}+\xi^{c_{g^{\prime},j}-c_{g^{\prime},i}}=0.$

{\it Case 4:} If $u_1= 0,~u_2>0$, and $i_{\pi_1(1)}\neq j_{\pi_1(1)}$ or $i_{\pi_{1}(m-k+\alpha)}\neq j_{\pi_{1}(m-k+\alpha)}$, we can find an array ${\bm c}^{\prime}={\bm c}+(q/2){\bm x}_{\pi_1(1)}\in G$ or ${\bm c}^{\prime}={\bm c}+(q/2){\bm x}_{\pi_{1}(m-k+\alpha)}$ for any array ${\bm c}\in G$. Similar to Case 1, we can obtain $\xi ^{c_{g,j}-c_{g,i}}+\xi ^{{c}^{\prime}_{g,j}-{c}^{\prime}_{g,i}}=0.$

{\it Case 5:} Suppose $u_1=0,~u_2>0$, $i_{\pi_1(1)}= j_{\pi_1(1)}$, and $i_{\pi_{1}(m-k+\alpha)}= j_{\pi_{1}(m-k+\alpha)}$ for all $\alpha = 1,2,\cdots,k$. Suppose that $\alpha^{\prime}$ is the largest non-negative integer satisfying $i_{\pi_{1}(m-k+\alpha^{\prime})}= j_{\pi_{1}(m-k+\alpha^{\prime})}=0$. Then we assume $\beta$ is the smallest integer which satisfies $i_{\pi_1(\beta)}\neq j_{\pi_1(\beta)}$. Here, we have $i_s=j_s=0$ for $s=\pi_{1}(m-k+\alpha^{\prime})+1,\pi_{1}(m-k+\alpha^{\prime})+2,\ldots,m-1$, and $s\neq \pi_{1}(m-k+\alpha)$ for $\alpha=\alpha^{\prime}+1,\alpha^{\prime}+2,\ldots,k$. Hence, it implies $\pi_{1}(\beta)<\pi_{1}(m-k+\alpha^{\prime})$ and $\pi_{1}(\beta-1)<\pi_{1}(m-k+\alpha^{\prime})$ according to the condition (C-3). Let $i^{\prime}$ and $j^{\prime}$ be integers that differ from $i$ and $j$, respectively, in the position $\pi_{1}(\beta-1)$. Similar to Case 2, we have
\begin{equation}
    \begin{aligned}
        \xi ^{c_{g,j}-c_{g,i}}+\xi ^{c_{g,j^{\prime}}-c_{g,i^{\prime}}}=0.
    \end{aligned}
\end{equation}

{\it Case 6:} Suppose $u_1= 0,~u_2>0$, $i_{\pi_1(1)}= j_{\pi_1(1)}$, and $i_{\pi_{1}(m-k+\alpha)}= j_{\pi_{1}(m-k+\alpha)}=1$ for all $\alpha = 1,2,\cdots,k$. Then we assume $\beta$ is the smallest integer which satisfies $i_{\pi_1(\beta)}\neq j_{\pi_1(\beta)}$. Since $i_s=j_s=0$ for $s=v+1,v+2,\cdots,m-k$ and $s\neq \pi_{1}(m-k+\alpha)$ for $\alpha=1,2,\ldots,k-1$, we can obtain $\pi_1(\beta)\leq v$ implying $\pi_1(\beta-1)\leq v$. Similar to Case 2, we have
\begin{equation}
    \begin{aligned}
        \xi ^{c_{g,j}-c_{g,i}}+\xi ^{c_{g,j^{\prime}}-c_{g,i^{\prime}}}=0.
    \end{aligned}
\end{equation}
    From Cases 1 to 6, the theorem can be proved.
\end{IEEEproof}
\begin{remark}
Taking $\sigma_2(l)=\pi_2(n-l+1)$ for $l=1,2,\ldots,n$ and $\pi_{1}(m-k+\alpha)=m-k+\alpha$ for $\alpha=1,2,\ldots,k$ in Theorem \ref{thm:GCAS_2}, (\ref{shen2022three_GBF}) can be represented as
\begin{equation}
\begin{aligned}
f=&\frac{q}{2}\left(\sum_{k=1}^{m-k-1}x_{\pi_1 (k)}x_{\pi_1 (k+1)}+\sum_{k=1}^{n-1}y_{\sigma_2 (k)}y_{\sigma_2 (k+1)}+x_m y_{\sigma_2 (1)}\right)+\sum_{l=1}^{m-k}\mu_l {\bm x}_{\pi_1(l)}{\bm x}_{m}\\ &+\sum_{l=1}^m p_l x_l+\sum_{s=1}^n\kappa_s y_s+p_0
\end{aligned}
\end{equation}
where $p_l,~\kappa_s\in \mathbb{Z}_q$. We can find that the result of Lemma \ref{lemma_shen2022three} is  a special case of Theorem \ref{thm:GCAS_2} by simply setting $k=1$, $\mu_{m-1}=\frac{q}{2}$, and $\mu_{l}=0$ for $l=1,\cdots,m-2$.
\end{remark}
\begin{eg}
Taking $q=2$, $m=5$, $n=2$, $k=2$, and $v=0$, we let $\pi_1=(1,2,4,3,5)$ and $\pi_2=(1,2)$. The generalized Boolean function is $f=x_1x_2+x_2x_4+y_1y_2+x_5y_1$ by setting $p_l,\kappa_s=0$. The array set $G$ is a GCAS of size $8$ when the truncated size $L_1=4$  $L_2=21$. We let $G=\{{\bm c}_0,{\bm c}_1,\cdots,{\bm c}_7\}$ and list the constituent arrays in Table \ref{table2}. Also, their AACF sum is shown as Fig. \ref{fig:GCAS2}.
\label{eg:GCAS2}
\end{eg}
\begin{table*}[ht]
\centering
\caption{The Constructed $(8,4,21)$-GCAS in Example \ref{eg:GCAS2}} \label{table2}
\begin{tabular}{|l|l|l|l|l|l|}
\hline
{$
{\bm c}_0=
\left( {\begin{array}{ccccccccccccccccccccccccccccccccccccccccccc}
0&	0&	0&	1&	0&	1&	1&	1&	0&	0&	1&	0&	1&	0&	1&	1&	0&	1&	1&	1&	0\\
0&	0&	0&	1&	0&	1&	1&	1&	0&	0&	1&	0&	1&	0&	1&	1&	0&	1&	1&	1&	0\\
0&	1&	0&	0&	0&	0&	1&	0&	0&	1&	1&	1&	1&	1&	1&	0&	0&	0&	1&	0&	0\\
1&	0&	1&	1&	1&	1&	0&	1&	1&	0&	0&	0&	0&	0&	0&	1&	1&	1&	0&	1&	1\\
\end{array}}
\right)
$}\\ \hline{$
{\bm c}_1=
\left( {\begin{array}{ccccccccccccccccccccccccccccccccccccccccccc}
0&	0&	0&	1&	1&	0&	0&	0&	0&	0&	1&	0&	0&	1&	0&	0&	0&	1&	1&	1&	1\\
0&	0&	0&	1&	1&	0&	0&	0&	0&	0&	1&	0&	0&	1&	0&	0&	0&	1&	1&	1&	1\\
0&	1&	0&	0&	1&	1&	0&	1&	0&	1&	1&	1&	0&	0&	0&	1&	0&	0&	1&	0&	1\\
1&	0&	1&	1&	0&	0&	1&	0&	1&	0&	0&	0&	1&	1&	1&	0&	1&	1&	0&	1&	0\\
\end{array}}
\right)
$}\\ \hline
{$
{\bm c}_2=
\left( {\begin{array}{ccccccccccccccccccccccccccccccccccccccccccc}
0&	0&	0&	1&	0&	1&	1&	1&	0&	0&	1&	0&	1&	0&	1&	1&	1&	0&	0&	0&	1\\
0&	0&	0&	1&	0&	1&	1&	1&	0&	0&	1&	0&	1&	0&	1&	1&	1&	0&	0&	0&	1\\
0&	1&	0&	0&	0&	0&	1&	0&	0&	1&	1&	1&	1&	1&	1&	0&	1&	1&	0&	1&	1\\
1&	0&	1&	1&	1&	1&	0&	1&	1&	0&	0&	0&	0&	0&	0&	1&	0&	0&	1&	0&	0\\
\end{array}}
\right)
$}\\ \hline {$
{\bm c}_3=
\left( {\begin{array}{ccccccccccccccccccccccccccccccccccccccccccc}
0&	0&	0&	1&	1&	0&	0&	0&	0&	0&	1&	0&	0&	1&	0&	0&	1&	0&	0&	0&	0\\
0&	0&	0&	1&	1&	0&	0&	0&	0&	0&	1&	0&	0&	1&	0&	0&	1&	0&	0&	0&	0\\
0&	1&	0&	0&	1&	1&	0&	1&	0&	1&	1&	1&	0&	0&	0&	1&	1&	1&	0&	1&	0\\
1&	0&	1&	1&	0&	0&	1&	0&	1&	0&	0&	0&	1&	1&	1&	0&	0&	0&	1&	0&	1\\
\end{array}}
\right)
$}
\\ \hline
{$
{\bm c}_4=
\left( {\begin{array}{ccccccccccccccccccccccccccccccccccccccccccc}
0&	0&	0&	1&	0&	1&	1&	1&	0&	0&	1&	0&	1&	0&	1&	1&	0&	1&	1&	1&	0\\
1&	1&	1&	0&	1&	0&	0&	0&	1&	1&	0&	1&	0&	1&	0&	0&	1&	0&	0&	0&	1\\
0&	1&	0&	0&	0&	0&	1&	0&	0&	1&	1&	1&	1&	1&	1&	0&	0&	0&	1&	0&	0\\
0&	1&	0&	0&	0&	0&	1&	0&	0&	1&	1&	1&	1&	1&	1&	0&	0&	0&	1&	0&	0\\
\end{array}}
\right)
$}\\ \hline
{$
{\bm c}_5=
\left( {\begin{array}{ccccccccccccccccccccccccccccccccccccccccccc}
0&	0&	0&	1&	1&	0&	0&	0&	0&	0&	1&	0&	0&	1&	0&	0&	0&	1&	1&	1&	1\\
1&	1&	1&	0&	0&	1&	1&	1&	1&	1&	0&	1&	1&	0&	1&	1&	1&	0&	0&	0&	0\\
0&	1&	0&	0&	1&	1&	0&	1&	0&	1&	1&	1&	0&	0&	0&	1&	0&	0&	1&	0&	1\\
0&	1&	0&	0&	1&	1&	0&	1&	0&	1&	1&	1&	0&	0&	0&	1&	0&	0&	1&	0&	1\\
\end{array}}
\right)
$}\\ \hline
{$
{\bm c}_6=
\left( {\begin{array}{ccccccccccccccccccccccccccccccccccccccccccc}
0&	0&	0&	1&	0&	1&	1&	1&	0&	0&	1&	0&	1&	0&	1&	1&	1&	0&	0&	0&	1\\
1&	1&	1&	0&	1&	0&	0&	0&	1&	1&	0&	1&	0&	1&	0&	0&	0&	1&	1&	1&	0\\
0&	1&	0&	0&	0&	0&	1&	0&	0&	1&	1&	1&	1&	1&	1&	0&	1&	1&	0&	1&	1\\
0&	1&	0&	0&	0&	0&	1&	0&	0&	1&	1&	1&	1&	1&	1&	0&	1&	1&	0&	1&	1\\
\end{array}}
\right)
$}\\ \hline
{$
{\bm c}_7=
\left( {\begin{array}{ccccccccccccccccccccccccccccccccccccccccccc}
0&	0&	0&	1&	1&	0&	0&	0&	0&	0&	1&	0&	0&	1&	0&	0&	1&	0&	0&	0&	0\\
1&	1&	1&	0&	0&	1&	1&	1&	1&	1&	0&	1&	1&	0&	1&	1&	0&	1&	1&	1&	1\\
0&	1&	0&	0&	1&	1&	0&	1&	0&	1&	1&	1&	0&	0&	0&	1&	1&	1&	0&	1&	0\\
0&	1&	0&	0&	1&	1&	0&	1&	0&	1&	1&	1&	0&	0&	0&	1&	1&	1&	0&	1&	0\\
\end{array}}
\right)
$}\\ \hline
\end{tabular}
\end{table*}

\section{Simulation Results}\label{sec:simulation}
In this section, we present the numerical results including the power radiation pattern and BER performance  by using our proposed 2D GCASs for massive MIMO systems with URA.

\subsection{Power Radiation Pattern}\label{sec:power_radiation_pattern}
\begin{figure}
    \centering
    \subfigure[GCAS-based precoding.]{
        \label{fig:GCAS1_a}
        \includegraphics[width=70mm]{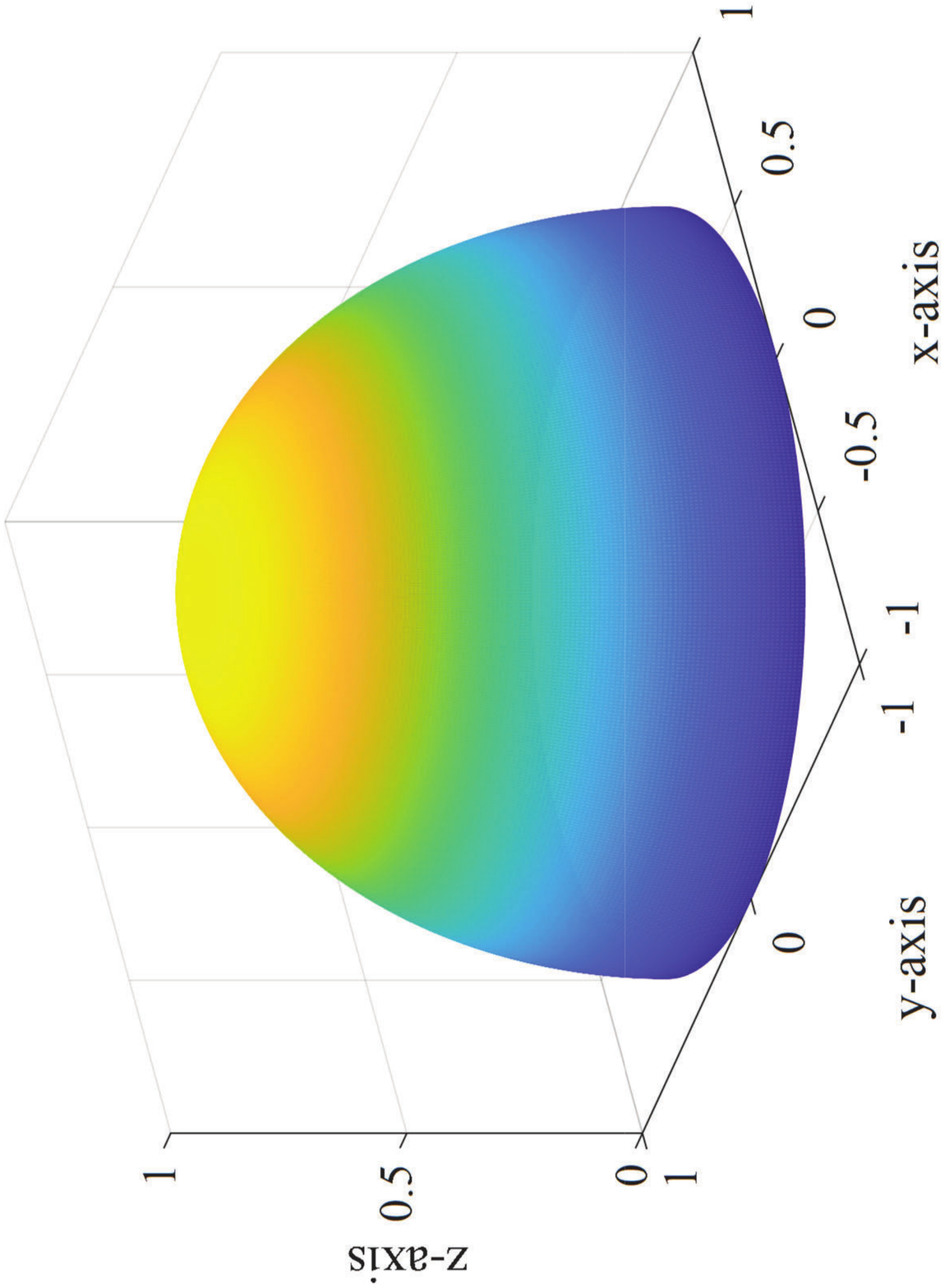}
    }\\
    \subfigure[ZC-based precoding.]{
        \label{fig:GCAS1_b}
        \includegraphics[width=70mm]{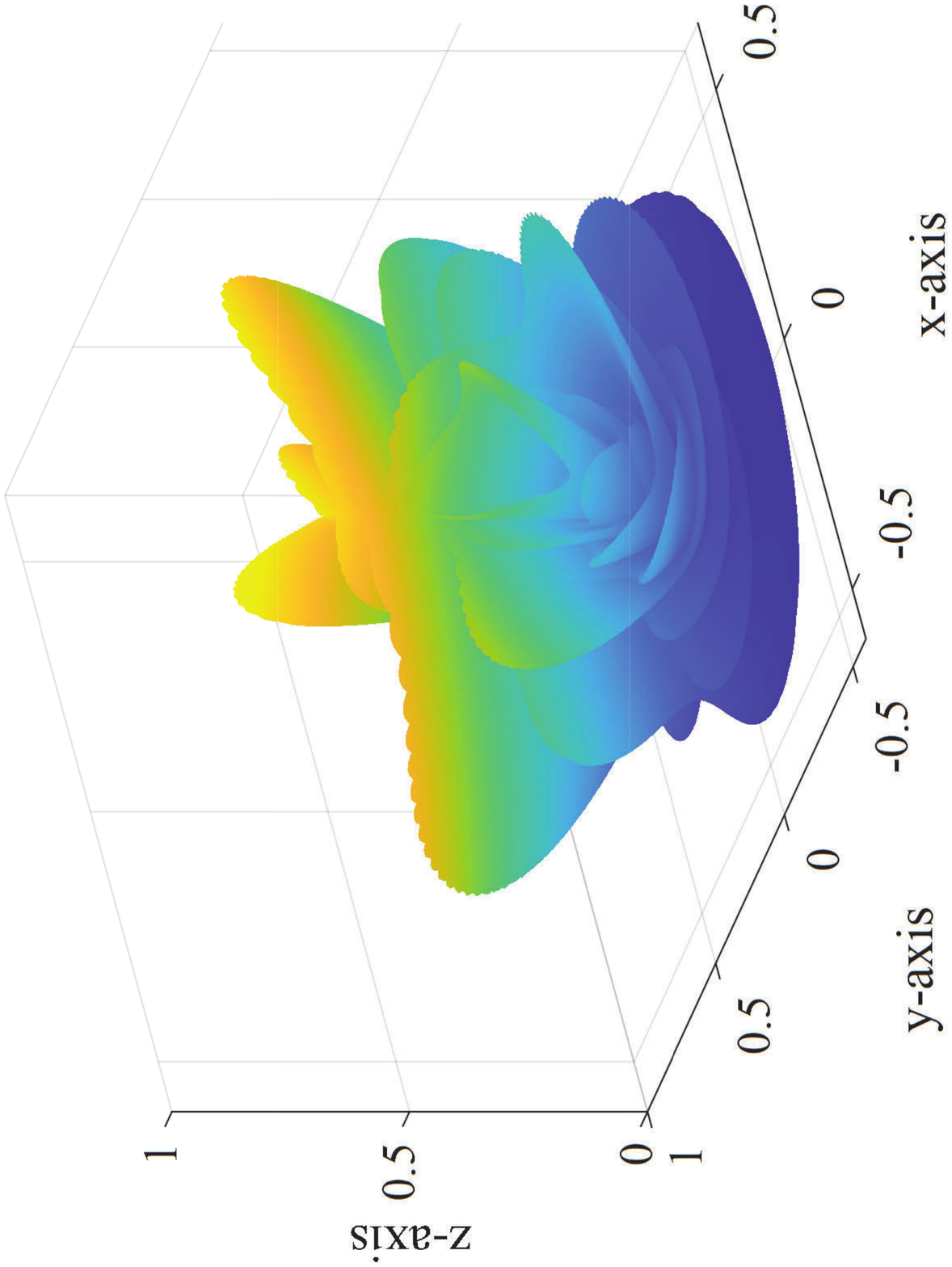}
    }
    \subfigure[Random-matrix-based precoding.]{
        \label{fig:GCAS1_c}
        \includegraphics[width=70mm]{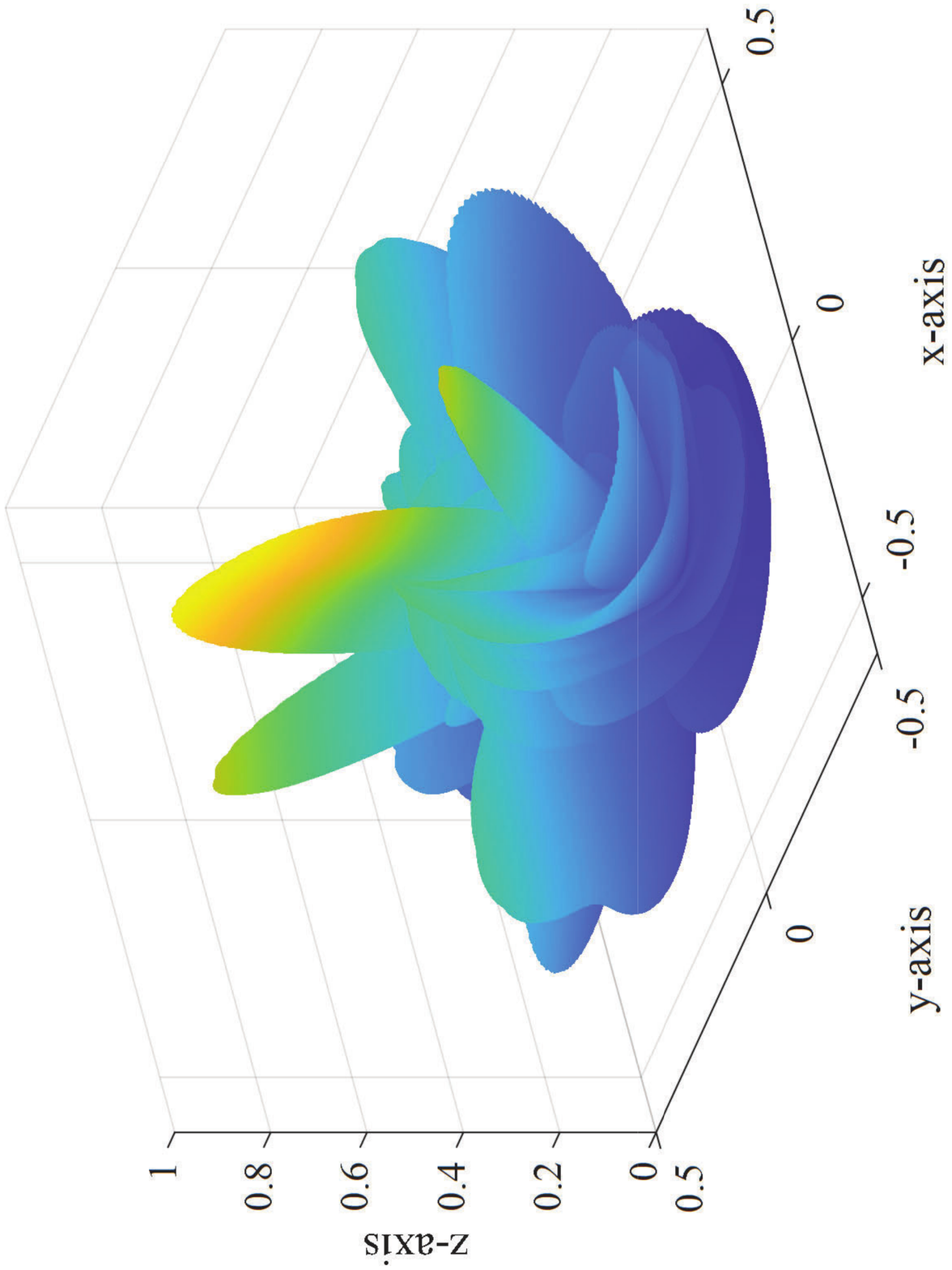}
    }
    \caption{Power radiation pattern with $4\times 33$ URA and $4\times 4$ STBC.}
\end{figure}

According to (\ref{received_singal}), the power radiation pattern $\sum_{n=0}^{N-1}\left|[\text{vec}({\bm A}(\varphi,\theta))^T\text{vec}({\bm W}_n)]\right|^2$ can be obtained. We first consider the massive MIMO system equipped with a URA of size $4\times 33$, i.e., $L_1=4$ and $L_2=33$. We take the GCS $G=\lbrace{\bm c}_0,{\bm c}_1,{\bm c}_2,{\bm c}_3\rbrace$ listed in Table \ref{table1} to generate the precoding matrices $\lbrace{\bm W}_0,{\bm W}_1,{\bm W}_2,{\bm W}_3\rbrace =\lbrace{(-1)^{{\bm c}_0}},{(-1)^{{\bm c}_1}},{(-1)^{{\bm c}_2}},{(-1)^{{\bm c}_3}}\rbrace$ with the omnidirectional property. The power radiation pattern of the GCAS-based scheme with array size $4\times 33$ is perfectly omnidirectional as illustrated in Fig \ref{fig:GCAS1_a}.

For the purpose of comparison, we also show the power radiation patterns of the precoding matrices based on Zadoff-Chu sequences and random-matrices whose elements are randomly generated from ``$+1$'' and ``$-1$''. The ZC-based precoder consists of four $4\times 33$ precoding matrices, which are obtained based on a ZC sequence of length $4$ and a ZC sequence of length $33$ \cite{li2021construction}. Fig. \ref{fig:GCAS1_b} illustrates the power radiation pattern of the ZC-based precoder. We can find that its power radiation pattern is not omnidirectional. The random-matrix-based precoder consists of four $4\times 33$ precoding matrices. The elements in the random-matrix-based precoding matrices are generated by selecting the elements from $\{1,-1\}$ with equal probability.  Fig. \ref{fig:GCAS1_c} describes the power radiation pattern of the random matrix-based precoder. We can observe that the power radiation pattern is not omnidirectional.
\begin{figure}
    \centering
    \subfigure[GCAS-based precoding.]{
        \label{fig:GCAS2_a}
        \includegraphics[width=70mm]{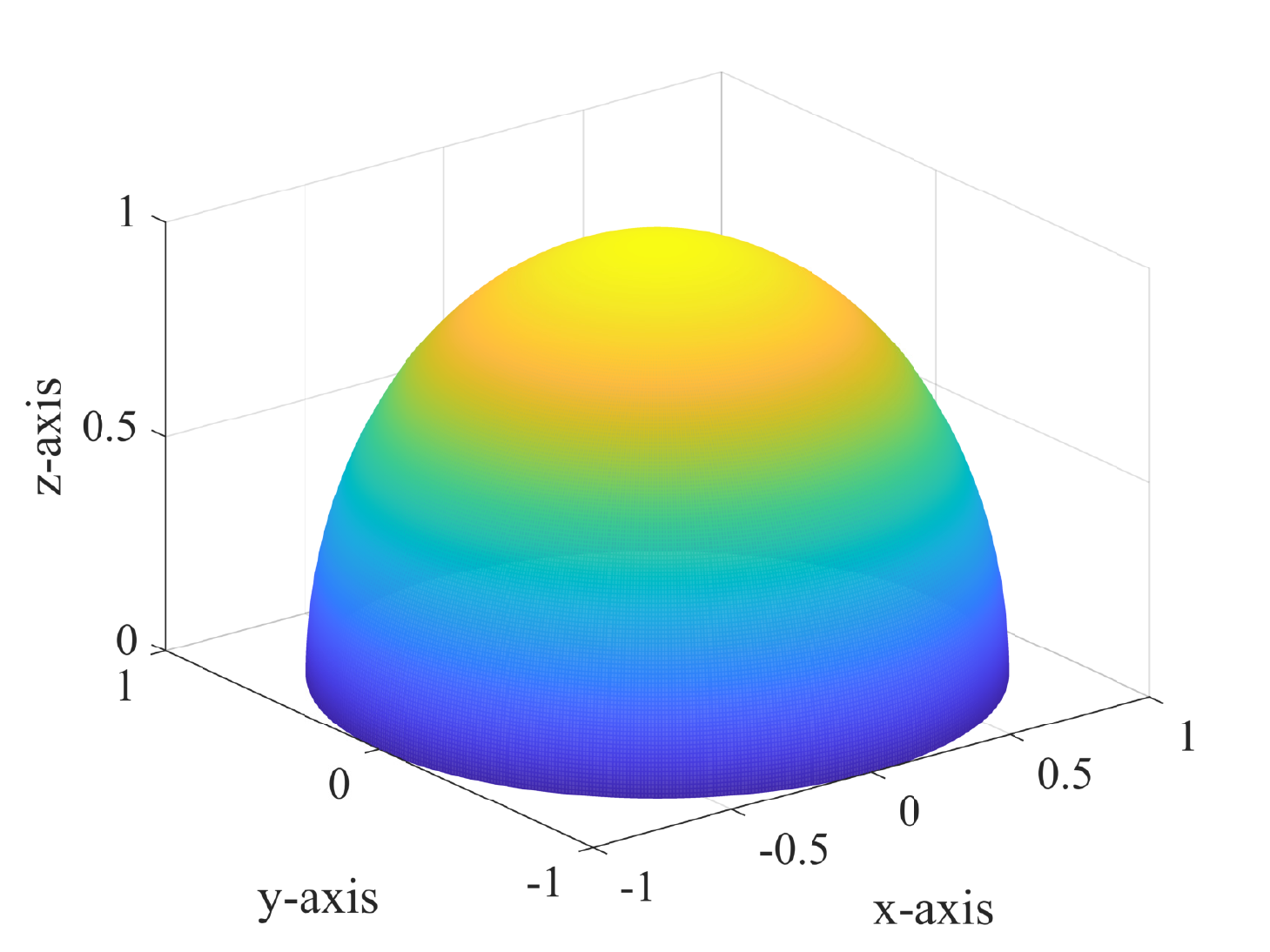}
    }\\
    \subfigure[ZC-based precoding.]{
        \label{fig:GCAS2_b}
        \includegraphics[width=70mm]{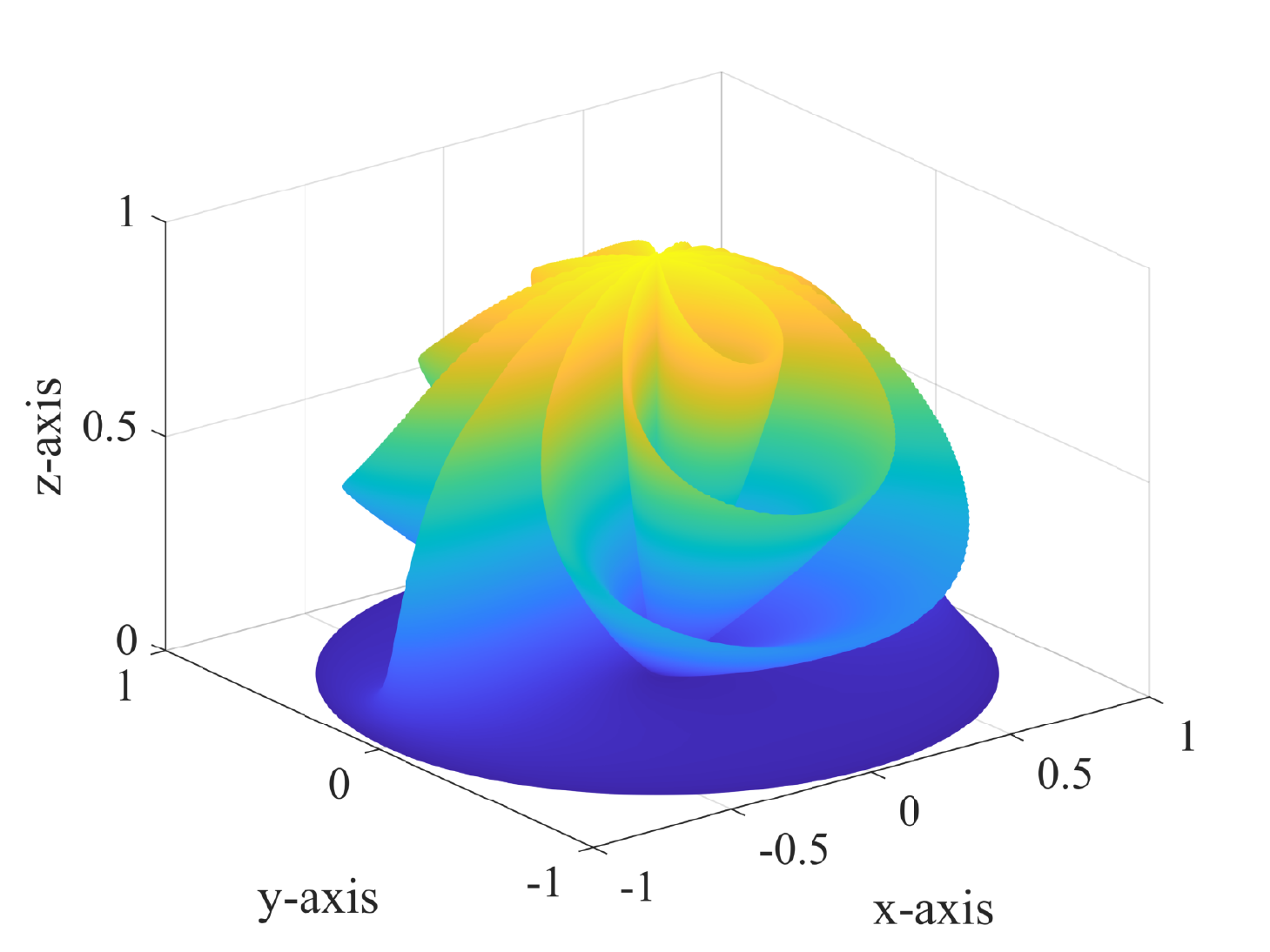}
    }
    \subfigure[Random-matrix-based precoding.]{
        \label{fig:GCAS2_c}
        \includegraphics[width=70mm]{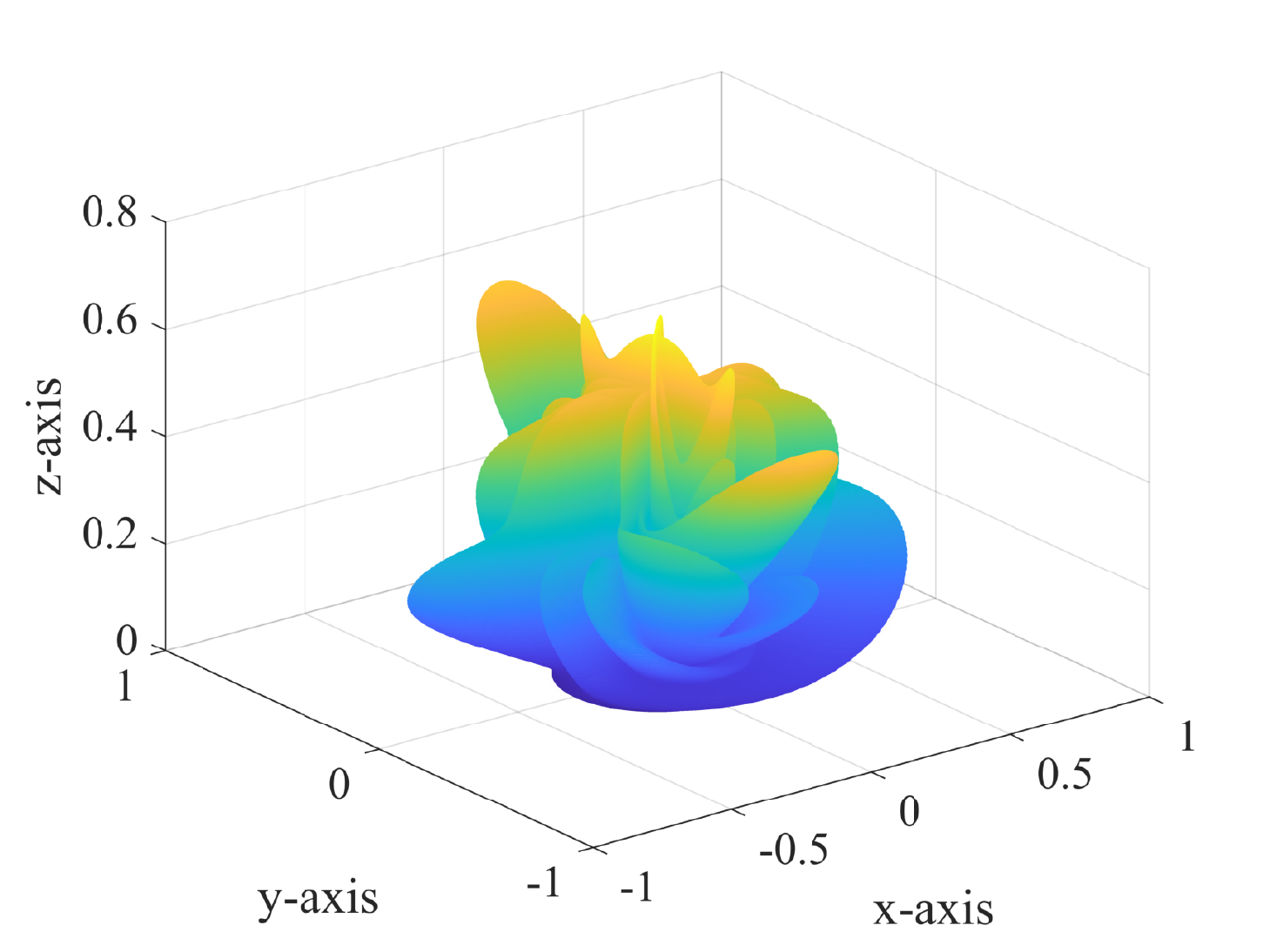}
    }
    \caption{Power radiation pattern with $4\times 21$ URA and $8\times 8$ STBC.}
\end{figure}

Next, we consider the massive MIMO system equipped with a URA of size $4\times 21$, i.e., $L_1=4$ and $L_2=21$. We use the GCS $G=\lbrace{\bm c}_0,{\bm c}_1,\cdots,{\bm c}_7\rbrace$ listed in Table \ref{table2} for the precoding matrix $\lbrace{\bm W}_0,{\bm W}_1,\cdots,{\bm W}_7\rbrace = \lbrace{(-1)^{{\bm c}_0}},{(-1)^{{\bm c}_1}}, \cdots,{(-1)^{{\bm c}_7}}\rbrace$. The power radiation pattern of the GCAS-based scheme with array size $4\times 21$ is described in \Fig\ref{fig:GCAS2_a}. The perfect omnidirectional property can be observed. We also see that the power radiation patterns of the ZC-based precoder and the random-matrix precoder shown in \Fig\ref{fig:GCAS2_b} and \Fig\ref{fig:GCAS2_c} are not omnidirectional. The ZC-based precoding matrices are obtained by a ZC sequence of length 4 and ZC sequence of 21 \cite{li2021construction}.

\subsection{Bit Error Rate Performance}
\begin{figure*}[!t]
\centering
\begin{center}
\extrarowheight=3pt
\includegraphics[width=120mm]{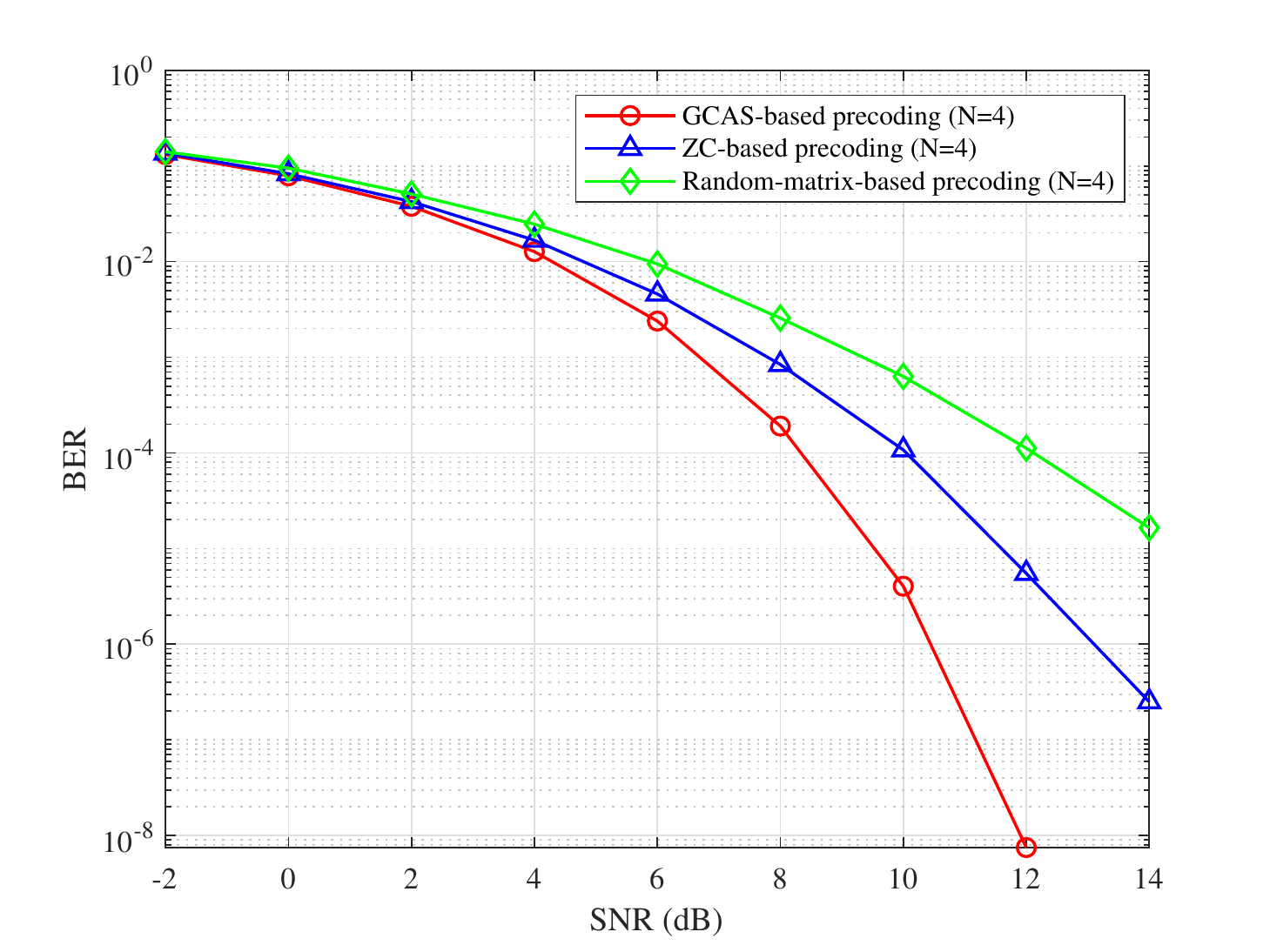}
\caption{BER performance of the different schemes for a $4\times33$ URA.}
\label{fig:GCAS15}
\end{center}
\end{figure*}

In this subsection, we present the BER performance of our proposed 2D GCAS-based schemes. We first consider the massive MIMO system equipped with a URA of size $4\times 33$. We let $N=4$ and then the $4\times 4$ orthogonal real STBC be presented as
\begin{equation}{\label{STBC}}
\begin{aligned}
{\bm S}= \left(\begin{matrix}s_0 & -s_1 & -s_2 & -s_3\\ s_1& s_0 & s_3 & -s_2\\ s_2&-s_3&s_0&s_1\\ s_3&s_2&-s_1&s_0 \end{matrix}\right),
\end{aligned}
\end{equation}
where $s_0,s_1,s_2,s_3$ are binary phase shift keying (BPSK) modulated symbols. We employ the maximum likelihood (ML) decoding here. For each realization, the elevation and the azimuth angles are uniformly distributed at random between $[0,\pi/2]$ and $[0, 2\pi]$, respectively. For comparison, the ZC-based precoder and random-matrix-based precoder are the same as mentioned in Section \ref{sec:power_radiation_pattern}. The BER performances of three different schemes are depicted in \Fig\ref{fig:GCAS15}. We can find that the 2D GCAS-based scheme outperform the others. At BER of $10^{-4}$, there are 1.6 dB and 3.6 dB gains over the ZC-based scheme and the random-matrix-based scheme, respectively.

\begin{figure*}[!t]
\centering
\begin{center}
\extrarowheight=3pt
\includegraphics[width=120mm]{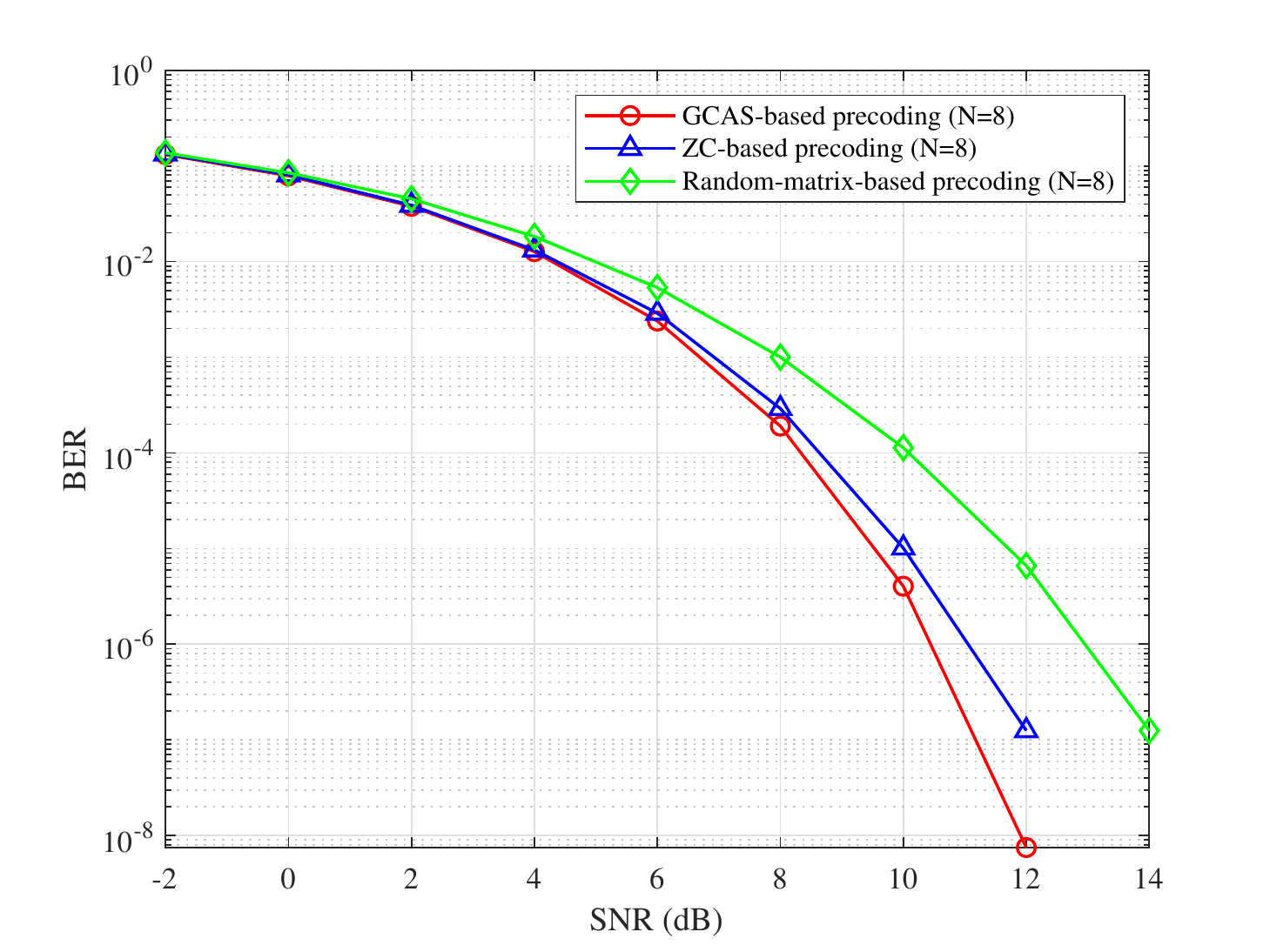}
\caption{BER performance of the different schemes for a $4\times21$ URA.}
\label{fig:GCAS16}
\end{center}
\end{figure*}

Next, we consider the massive MIMO system equipped with a URA of size $4\times 21$. We consider $8\times 8$ STBC and the $8\times 8$ orthogonal real STBC is given by
\begin{equation}{\label{STBC_8}}
\begin{aligned}
{\bm S}= \left(\begin{matrix}
s_0 & s_1 & s_2 & s_3 &s_4 & s_5 & s_6 & s_7\\
-s_1 & s_0 & s_3 & -s_2 &s_5 & -s_4 & -s_7 & s_6\\
-s_2 & -s_3 & s_0 & s_1 &s_6 & s_7 & -s_4 & -s_5\\
-s_3 & s_2 & -s_1 & s_0 &s_7 & -s_6 & s_5 & -s_4\\
-s_4 & -s_5 & -s_6 & -s_7 &s_0 & s_1 & s_2 & s_3\\
-s_5 & s_4 & -s_7 & s_6 &-s_1 & s_0 &-s_3 & s_2\\
-s_6 & s_7 & s_4 & -s_5 &-s_2 & s_3 & s_0 & -s_1\\
-s_7 & -s_6 & s_5 & -s_4 &-s_3 & s_2 & s_1& s_0\\\end{matrix}\right)
\end{aligned}
\end{equation}
where $s_0,s_1,\cdots,s_7$ are BPSK modulated symbols. We also take the ZC-based precoding and random-matrix-based precoding for comparison. The BER performance comparison for these three different schemes is depicted in Fig. \ref{fig:GCAS16}. At BER of $10^{-4}$, there are 0.2 dB and 1.8 dB gains over the ZC-based scheme and the random-matrix-based scheme, respectively. As a result, the 2D GCASs are good candidates as precoding matrices for omnidirectional transmission in massive MIMO systems.

\section{Conclusion}\label{sec:conclusion}
In this paper, constructions of 2D GCASs with flexible array sizes have been proposed in Theorems \ref{thm:GCAS} and \ref{thm:GCAS_2}. Our constructions can be obtained directly from 2D GBFs without the aid of special sequences. Besides, our  proposed GCASs have flexible array sizes which can fit more antenna configuration. Furthermore, Theorem 2 can include the results in \cite{shen2022three} as a special case. Simulation results showed that the omnidirectional transmission can be achieved when the precoding matrices are based on the proposed GCASs. The BER performance due to their omnidirectional power radiation patterns, the ZC-based scheme and random-matix-based have inferior performances because their power radiation patterns both are not ideally omnidirectional.  Although Theorems \ref{thm:GCAS} and \ref{thm:GCAS_2} can provide direct constructions of 2D GCASs, the first dimension has size $L_1$ limited to $2^n$. Therefore, the future work includes the extension of constructions of 2D GCASs of which both dimensions have non-power-of-two sizes.
\bibliographystyle{IEEEtran}
\bibliography{IEEEabrv,GCAS}

\end{document}